\documentclass[11pt]{article}
\usepackage{jheppub}
\usepackage{amsmath,amssymb,amsfonts,graphicx,slashed,color}
\usepackage{subcaption}
\usepackage{tikz}
\usepackage{tikz-3dplot}
\usepackage[utf8]{inputenc}
\usepackage{pgfplots}
\usepgfplotslibrary{colormaps}
\usetikzlibrary{decorations.markings}
\usetikzlibrary{decorations.pathmorphing}
\usepgfplotslibrary{fillbetween}
\usetikzlibrary{patterns}
\usetikzlibrary{shapes.misc}

\newcommand{\be}{\begin{equation}}
\newcommand{\ee}{\end{equation}}
\newcommand{\bea}{\begin{eqnarray}}
\newcommand{\eea}{\end{eqnarray}}
\newcommand{\beq}{\begin{equation}}
\newcommand{\eeq}{\end{equation}}
\newcommand{\beqn}{\begin{eqnarray}}
\newcommand{\eeqn}{\end{eqnarray}}

\usepackage{upgreek}

\title{Flat space physics from AdS/CFT }

\author{Eliot Hijano ${}^{\Diamond}$}

\affiliation{${}^{\Diamond}$ Department of Physics and Astronomy, University of British Columbia,\\
6224 Agricultural Road, Vancouver, B.C.\ V6T 1Z1, Canada.}

\emailAdd{eliothijanoc@gmail.com}

\abstract{
We propose a formula relating scattering S-matrix amplitudes to correlators of a conformal field theory. The proposal implements a flat limit of the field theory, providing an indirect microscopic description of gravitational theories with asymptotically flat boundary conditions. The formula is valid for both massive and massless external particles, and reduces to existing expressions in the literature when all particles are either simultaneously massless or massive. We test the result in various $(2+1)$-dimensional examples such as simple BMS$_3$ invariant correlators and blocks. We also study two-point correlators in conformal field theory deficit states to obtain known expressions for non-trivial scattering in asymptotically flat conical geometries. 
}

\keywords{}

\begin{document}

\tikzset{->-/.style={decoration={
  markings,
  mark=at position #1 with {\arrow{>}}},postaction={decorate}}}

\def \L {10}
\def \H {1.5*\L}

\tikzset{
    mark position/.style args={#1(#2)}{
        postaction={
            decorate,
            decoration={
                markings,
                mark=at position #1 with \coordinate (#2);
            }
        }
    }
}

\tikzset{
  pics/carc/.style args={#1:#2:#3}{
    code={
      \draw[pic actions] (#1:#3) arc(#1:#2:#3);
    }
  }
}

\tikzset{point/.style={insert path={ node[scale=2.5*sqrt(\pgflinewidth)]{.} }}}

\tikzset{->-/.style={decoration={
  markings,
  mark=at position #1 with {\arrow{>}}},postaction={decorate}}}

  \tikzset{-dot-/.style={decoration={
  markings,
  mark=at position #1 with {\fill[red] circle [radius=3pt,red];}},postaction={decorate}}} 

 \tikzset{-dot2-/.style={decoration={
  markings,
  mark=at position #1 with {\fill[blue] circle [radius=3pt,blue];}},postaction={decorate}}} 

    \definecolor{darkgreen}{RGB}{0,180,0}

    \definecolor{purple2}{RGB}{222,0,255}

 \tikzset{-dot3-/.style={decoration={
  markings,
  mark=at position #1 with {\fill[purple2] circle [radius=3pt,purple2];}},postaction={decorate}}} 

\tikzset{snake it/.style={decorate, decoration=snake}}

    \tikzset{cross/.style={cross out, draw=black, minimum size=2*(#1-\pgflinewidth), inner sep=0pt, outer sep=0pt},
cross/.default={3pt}}


\maketitle

\parskip=10pt


\section{Introduction}
The holographic principle  \cite{tHooft:1999rgb,Susskind:1994vu} relates a theory of quantum gravity in a volume of space-time to a non-gravitational theory in a lower dimensional boundary. A concrete example for this principle is the AdS/CFT correspondence \cite{Maldacena:1997re}, relating a theory of quantum gravity with anti-de Sitter (AdS) asymptotics to a conformal field theory (CFT).  Generalizing this duality to other asymptotics is of great relevance, as it would allow for a holographic understanding of physics in a Universe similar to our own. In this note, we study physics in flat space-time from an indirect holographic perspective.

Formulating holography in asymptotically flat space-times entails establishing a microscopic theory that naturally possesses all holographic degrees of freedom at the boundary. Once the microscopics are understood, it is in principle possible to compare the observables in the gravitational theory with the ones at the boundary. Alternatively, in this note we will seek to describe holography in flat space through a flat limit of the well established AdS/CFT conjecture. The observables in the gravitational theory can then be understood as particular limits of CFT observables, providing an indirect microscopic description of the gravitational theory. 

The observables we will be concerned with in this paper are the S-matrix amplitudes of the asymptotically flat geometry, which we will study as flat limits of CFT correlators. The general philosophy we  explore  is not new. The relation between the flat space S-matrix and the AdS/CFT correspondence has been discussed extensively in the literature \cite{Giddings:1999jq,Gary:2009mi,Giddings:1999qu,Balasubramanian:1999ri,Penedones:2010ue,Fitzpatrick:2011jn}. See also \cite{Maldacena:2011nz,Raju:2012zr} for an approach relating momentum space CFT correlators to S-matrices in flat space. More recently, it has been understood how analyticity and unitarity of the S-matrix follows from the structure of a holographic conformal field theory \cite{Fitzpatrick:2011hu,Fitzpatrick:2011dm}.  In \cite{Paulos:2016fap}, an expression for the flat space S-matrix that involves external massive particles was proposed and used to implement the bootstrap programme. The approach presented in this paper will clarify the connection between the flat limit presented in \cite{Paulos:2016fap} involving massive particles and the results presented in \cite{Gary:2009ae,Penedones:2010ue,Fitzpatrick:2011hu,Fitzpatrick:2011dm} that apply to massless particles.  The formula we propose in this work is obtained by applying the following strategy. We first define a region of an asymptotically AdS space-time that turns into an asymptotically flat geometry upon the implementation of a flat limit. In such region, we define local operators from the perspective of the conformal field theory by applying the HKLL proposal \cite{Hamilton:2006az}. Finally, we perform a Fourier transform on the coordinates of the flat region such that the resulting operator is defined at a particular value of momentum. Correlators of such operators in the large radius of AdS limit are then related to the S-matrices of flat space, resulting in a simple formula that we advertise here
\be\label{eq:advertise}
{\cal S}\{ p_i  \} =   \lim_{l\rightarrow\infty} l^{{d-3}\over 2} \left[ \prod_i   C(p_i) \int dt_i \, e^{\pm i \omega_i t_i}\right] \langle 0 \vert  {\cal O}\left(   \tau_1   ,\Omega_1 \right)   \cdots      \vert 0 \rangle\, .
\ee
The parameter $l$ stands for the AdS radius, and the normalization constants $C(p_i)$ can be found in formula \ref{eq:normalization} in the main text. The operators inside the CFT correlator have conformal dimensions $\Delta_i\sim{\cal O}(1)$ if the resulting particles in flat space are massless, or otherwise obey $\Delta_i=m_i l$ if the resulting particles have mass $m_i$. The operators in the correlator are inserted at momenta-dependent locations given by
\be
\tau_i =\pm{{\pi}\over 2} \pm i \cosh^{-1}{{\omega_i}\over{k_i}}   +{{t_i}\over l}   ,\quad \text{and}\quad  \Omega_i-\chi_i ={{\pi}\over 2}  \mp        {{\pi}\over 2} \, .
\ee
Here, the Lorentz vector $p_i=(\omega_i, \mathbf{k}_i)$ determines energy-momentum in cylindrical coordinates, where $k_i=\vert \mathbf{k}_i \vert$ and $\chi_i$ parametrizes a point on the unit sphere. The sign choice is related to the choice between ``in" and ``out" states in the resulting S-matrix.

As an application of the proposal \ref{eq:advertise}, we will study the scattering of particles against an asymptotically flat cone geometry. A  non-relativistic study can be found in \cite{Deser:1988qn,tHooft:1988qqn}, and a more recent relativistic calculation can be found in \cite{Spinally:2000ii}. The S-matrix amplitude obtained in these works will be matched with the flat limit of CFT$_2$ correlators in a conical defect state, suggesting that non-trivial CFT correlators can be used to study non-trivial scattering processes in asymptotically flat space-times. 

Holography in flat space-time should also be understood independently of AdS/CFT. The asymptotic structure of flat-space suggests that a theory with BMS symmetry should arise as a dual to a gravitational theory with Minkowski asymptotics \cite{Sachs:1962wk}. In this note, we will revisit the construction of observables in such a theory in $(2+1)$ dimensions. We will construct operators associated to irreducible unitary representations of the symmetry algebra \cite{Campoleoni:2016vsh}, and construct simple correlators fixed by the symmetry. The results obtained this way will be then matched to the flat limits we construct in the main text.

It is also worth mentioning the relation between the results presented here and our previous work concerning holographic correlators in the context of $(2+1)$-dimensional flat holography \cite{Hijano:2017eii,Hijano:2018nhq}. In those papers, we studied correlators of operators associated to highest weight representations of the BMS$_3$ symmetry group. Such correlators were mapped to holographic structures consisting on Feynman diagrams integrated over null lines falling from the boundary at the locations of the insertions of the BMS$_3$ primary operators. Highest weight representations are however explicitly non-unitary \cite{Campoleoni:2016vsh,Bagchi:2019unf,Araujo:2018dem}, and thus unphysical in the context of holography. Indeed, one can regard the non-local nature of the holographic construction as a symptom of the choice of representations. It is of immediate interest to consider the unitary representations of the BMS$_3$ group. These representations naturally lead to operators in momentum space, and thus correlators are simply the S-matrix elements of the bulk theory, which are the observables we study throughout this work.

The paper is organized as follows; In section \ref{sec:SMATRIX} we assemble a proposal for the S-matrix scattering amplitude in an asymptotically flat space-time from conformal correlators. We test the resulting formula by computing Poincar\'e invariants from simple vacuum CFT correlators. In section \ref{sec:BMS3} we review $(2+1)$-dimensional gravity in flat space and its relation to the BMS$_3$ group. We review the definition of unitary representations of the BMS$_3$ algebra, introduce BMS operators, and construct simple correlators that match the flat limits in the previous section.  Section \ref{sec:CONE} is devoted to the study of light particles propagating around a conical deficit/cosmic string in flat space, from the indirect holographic perspective proposed above. We close the paper with some discussion and future work in section \ref{sec:DISC}.


\section{Flat space scattering amplitudes from CFT correlators}\label{sec:SMATRIX}
Ideally, in a holographic theory of gravity in flat space, scattering amplitudes would be obtained from specific quantum theories realizing the holographic principle. However, as of now, there are no examples of candidate theories with BMS$_3$ invariance and large central charge\footnote{Explicit large $c$ conformal field theories dual to Einstein gravity in $2+1$ dimensions are also unknown  \cite{Witten:2007kt,Castro:2011zq}, so this issue is not a feature contingent to flat space-times. Interesting work concerning duals to flat space in three dimensions include \cite{Barnich:2012rz,Barnich:2013yka}.}. An alternative strategy to provide flat space with some microscopic understanding is to regard flat space-time as a flat limit of anti-de Sitter space, and use the AdS/CFT duality to indirectly relate observables in the gravitational theory with observables in a theory without gravity. Such a strategy might prove useful to probe fundamental questions like the resolution of the information loss problem in black holes with flat  asymptotics.

The purpose of this section is to motivate the introduction of a map between CFT correlators and scattering amplitudes in asymptotically flat space-times. The construction presented in this section reduces to the prescription used in \cite{Fitzpatrick:2011hu,Fitzpatrick:2011dm} when the external particles in the S-matrix are massless, and is equivalent to the map proposed in \cite{Paulos:2016fap} when the external particles are massive. It is worth mentioning that the proposal we develop in this section is valid for any dimension, even though the tests we perform throughout the paper focus on the ($2+1$)-dimensional case. 

The strategy is quite simple; We first define a region of AdS that turns into flat space upon the implementation of a flat limit. The limit is realized by taking the AdS length scale to infinity. Inside this region, we then define local bulk operators using the HKLL prescription \cite{Hamilton:2006az}. In order to define S-matrix amplitudes, these operators must be taken to momentum space through a Fourier transform. This construction yields a smearing of CFT primaries, whose correlators turn into S-matrix amplitudes upon taking the flat limit. 

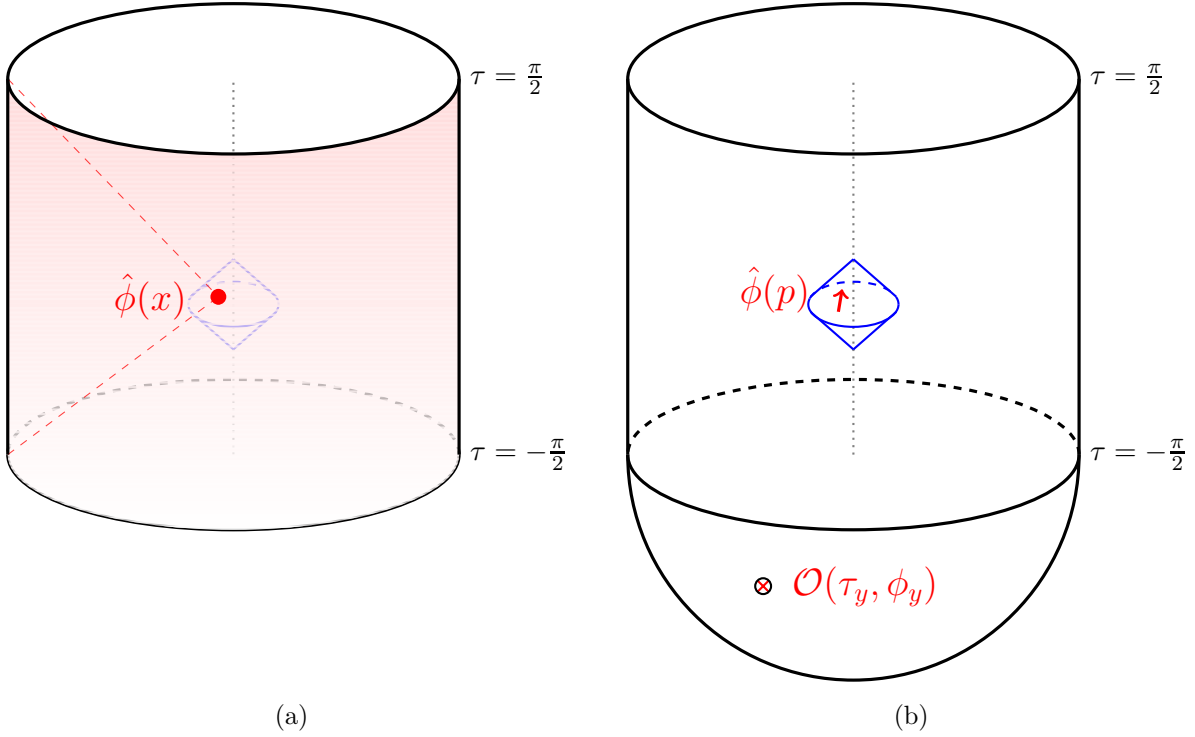
\begin{figure}[]
\centering
\begin{tabular}{cc}
\begin{subfigure}[t]{0.49\textwidth}
\centering
\begin{tikzpicture}
\draw[white,very thick] (5,2) arc (0:-180:3 and 3);
\draw[white,very thick,name path=TOP](1.25,7) arc (180:360:0.75 and 0.25);
\draw[gray, thick,dotted] (2,2) -- (2,7) node [pos=1,above=2]{{$ $}} node [pos=0,below=2]{{$ $}};
\draw[black,very thick,dashed](5,2) arc (0:180:3 and 1);
\draw[black,very thick](-1,2) arc (180:360:3 and 1);
\draw[blue,thick](2.6-0.03,4+0.1)--(2,4.6);
\draw[blue,thick](1.4+0.03,4+0.1)--(2,4.6);
\draw[blue,thick,dashed](2.6,4) arc (0:180:0.6 and 0.3);
\draw[blue,thick](2.6,4) arc (0:-195:0.6 and 0.3);

\draw[blue,thick](2.6-0.03,4-0.1)--(2,3.4);
\draw[blue,thick](1.4+0.03,4-0.1)--(2,3.4);
\shade[top color=red!15!white,opacity=0.75]  (-1,2) arc (180:360:3 and 1)   --   (5,7) arc (0:-180:3 and 1)    -- cycle;
\draw[red,dashed,opacity=0.75] (-1,2) -- (2-0.2,4.1);
\draw[red,dashed,opacity=0.75] (-1,7) -- (2-0.2,4.1);
\draw[red,very thick,-dot-=1] (2-0.2,4)--(2-0.2,4.1) node[pos=1,left=8]{\Large $\hat{\phi}(x)$};
\draw[black,very thick] (-1,2) -- (-1,7);
\draw[black,very thick] (5,2) -- (5,7) node [pos=1,right]{$\tau={{\pi}\over 2}$} node [pos=0,right]{$\tau=-{{\pi}\over 2}$};
\draw[black,very thick](2,7) ellipse (3 and 1);
\end{tikzpicture}
\caption{}
\end{subfigure}\hspace{2mm}
&
\begin{subfigure}[t]{0.49\textwidth}
\centering
\begin{tikzpicture}
\draw[white,very thick,name path=TOP](1.25,7) arc (180:360:0.75 and 0.25);
\draw[gray, thick,dotted] (2,2) -- (2,7) node [pos=1,above=2]{{$ $}} node [pos=0,below=2]{{$ $}};
\draw[black,very thick,dashed](5,2) arc (0:180:3 and 1);
\draw[black,very thick](-1,2) arc (180:360:3 and 1);
\draw[blue,thick](2.6-0.03,4+0.1)--(2,4.6);
\draw[blue,thick](1.4+0.03,4+0.1)--(2,4.6);
\draw[blue,thick,dashed](2.6,4) arc (0:180:0.6 and 0.3);
\draw[blue,thick](2.6,4) arc (0:-195:0.6 and 0.3);

\draw[blue,thick](2.6-0.03,4-0.1)--(2,3.4);
\draw[blue,thick](1.4+0.03,4-0.1)--(2,3.4);
\draw[red,very thick,->] (2-0.2,4-0.1)--(2-0.15,4.2) node[pos=1,left=8]{\Large $\hat{\phi}(p)$};
\draw[black,very thick] (-1,2) -- (-1,7);
\draw[black,very thick] (5,2) -- (5,7) node [pos=1,right]{$\tau={{\pi}\over 2}$} node [pos=0,right]{$\tau=-{{\pi}\over 2}$};
\draw[black,very thick](2,7) ellipse (3 and 1);
\draw[black,very thick] (5,2) arc (0:-180:3 and 3);
\draw[thick] (0.8,0.25) node[cross,red] {};
\draw[thick] (0.8,0.25) circle (3pt) node[right]{\Large \, {\color{red} ${\cal O}(\tau_y,\phi_y)$}};
\end{tikzpicture}
\caption{}
\end{subfigure}
\end{tabular}
    \caption{Construction of flat space scattering states from states in a conformal field theory. \textbf{a)} The bulk field $\hat{\phi}(x)$ is placed inside a scattering region around the center of global AdS (blue). The local operator can be reconstructed semi-classically in the boundary using the HKLL formula \ref{eq:HKLL}. The reconstruction involves  the shaded red part of the boundary, which is  space-like separated to the scattering region.   \textbf{b)} After performing a Fourier transform with respect to the flat space coordinates and implementing a flat limit, the reconstruction is performed by placing a primary operator in a Euclidean continuation of the boundary manifold. If the flat space particle is massless, the CFT operator is inserted at a real value of global time $\tau=\pm \pi/2$. 
}\label{fig:SmatrixMap}
\end{figure}

We start by defining a flat space scatteting region inside the bulk of AdS. The metric of the bulk geometry reads
\be
ds^2 = {{l^2 }\over{ \cos^2\rho}}\left(   -d\tau^2 +d\rho^2+\sin^2\rho \, d\Omega_{d-1}^2        \right)\, .
\ee
The  following coordinate replacement turns this line element into that of Minkowski space in cylindrical coordinates upon the limit $l\rightarrow \infty$
\be\label{eq:FlatRegion}
\tau = {t\over l}\, , \quad \text{and}\quad \rho = {r\over l}\, .
\ee
The region defined by these coordinates has been drawn in blue in figure \ref{fig:SmatrixMap}.a). 

Before we take a large $l$ limit, we define a local bulk operator $\hat{\phi}(x)$ at some point $x$ inside the scattering region specified in \ref{eq:FlatRegion}. The expression for such field as a function of CFT operators reads, in the large $N$ limit \cite{Hamilton:2006az}
\be\label{eq:HKLL}
\hat{\phi}^{(0)}(x) =\int_{\text{space-like}} d^d y \, K_{\Delta}(x\vert y) {\cal O}(y)\, .
\ee
The physical bulk field away from the semi-classical limit involves multi-trace corrections \cite{Kabat:2011rz,Kabat:2016zzr}, which we will not consider throughout this work. The mass of the bulk field is related to the conformal dimension of the primary operator as $m^2 l^2=\Delta(\Delta-d)$. This means that in order to study massive particles in flat space, we will consider primary operators with conformal dimensions $\Delta=m l$ in the large $l$ limit. The case of massless particles corresponds to primaries with conformal dimensions that do not scale with $l$.

The integral in equation \ref{eq:HKLL} is performed over the region of the AdS boundary that is space-like separated from the bulk point $x$. For the case at hand, in which the point $x$ is inside the flat space scattering region defined above, the integration region is approximately the global cylinder from $\tau=-\pi/2$ to $\tau=\pi/2$. This region is shaded in red in figure \ref{fig:SmatrixMap}.a). 

We are interested in the construction of  S-matrix amplitudes, so we need to study operators in momentum space. More specifically, we are interested in the momentum space dual to the coordinates $x$ inside the scattering region \ref{eq:FlatRegion}. We thus perform a Fourier transform with respect to such coordinates
\be\label{eq:Ppv1}
\hat{\phi}(p)=\int d^d x \, e^{i p\cdot x}\hat{\phi}(x) =\int_{\text{space-like}} d^d y \left[ \int d^d x \, e^{i p\cdot x}   K_{\Delta}(x\vert y)\right] {\cal O}(y) \, .
\ee
Note that the S-matrix amplitudes we seek to construct are objects defined in the asymptotic null boundary of flat space, and so the Fourier transform against plane-wave solutions presented here is valid as long as the geometry inside the scattering region \ref{eq:FlatRegion} is asyptotically flat under the large AdS radius limit. The expression between brackets in formula \ref{eq:Ppv1} can be simplified greatly in this limit. The kernel must obey the scalar field equation with respect to the coordinate $x$. This means that a general solution for the kernel can be written as
\be\label{eq:Kpp}
 K_{\Delta}(x\vert y) = \int d^{d+1} p' e^{-i p' \cdot x} k_{p'}(y)= \int d^{d+1} p' e^{i k' \left( {{\omega'}\over{k'}}  t_x -r_x \cos(\Omega_x-\chi')          \right)} k_{p'}(y)\, ,
\ee
It is also true that kernel must be invariant under AdS isometries, and so it can only depend on the AdS invariant distance
\be
\sigma(x\vert y)  = {{\cos(\tau_x-\tau_y)-\sin\rho_x\sin\rho_y \cos(\Omega_x-\Omega_y)}\over{\cos\rho_x\cos\rho_y}} \, .
\ee
placing $x$ inside the scattering region, and expanding the location of the boundary point as
\be
\rho_y={{\pi}\over2}\, , \quad \text{and}\quad \tau_y=\tau^{(0)}_y + {{t_y}\over{l}}\, ,
\ee
we obtain
\be\label{eq:sigma}
\sigma(x\vert y)  \cos\rho_y =   \cos\tau_y^{(0)}+{1\over{l}}\left( t_x \sin\tau_y^{(0)}  -r_x\cos(\Omega_x-\Omega_y)- t_y \sin\tau_y^{(0)}   \right)+{\cal O}(l)^{-2}\, .
\ee
This is consistent with formula \ref{eq:Kpp} if
\be\label{eq:TPO}
\begin{split}
\sin \tau_y^{(0)}&=\pm{{\omega'}\over{k'}} \, , \quad \text{and}\quad 
\chi'-\Omega_y= {{\pi}\over 2}  \mp        {{\pi}\over 2}  \, .\\
\end{split}
\ee
Note that the different signs here can be exchanged if we continue the momentum vector to its negative value $p'\rightarrow -p'$. We can thus understand the choice of sign in \ref{eq:TPO} as a choice between ``in'' and ``out'' states. We conclude that the Fourier transform of the kernel $ K_{\Delta}(x\vert y)$ constrains the boundary point $y$ to a fringe of global time parametrized by $t_y$ around a particular point in the CFT, specified by formulas \ref{eq:TPO}. Note that physical momenta have $\omega'\geq k'$, and so the value of $\tau_y^{(0)}$ obtained here is complex. Indeed, one can think of the point $\tau_y^{(0)}$ as a point in a euclidean half sphere attached to the Lorentzian geometry at $\tau=\pm\pi/2$ depending on the sign choice in formulas \ref{eq:TPO}. This can be seen explicitly by replacing
\be
\tau_y^{(0)}=\pm{{\pi}\over 2} + i \tilde{\tau}_y\, ,
\ee
where $\tilde{\tau}_y$ is now a coordinate in the Euclidean half sphere,  as formulas \ref{eq:TPO} now imply a real solution for $\tilde{\tau}_y$
\be
 \tilde{\tau}_y=\pm \cosh^{-1}{{\omega'}\over{k'}}\, .
\ee
In figure \ref{fig:SmatrixMap}.b) we have drawn a picture representing the insertion of the CFT primaries in the Euclidean half sphere, and the resulting bulk operator in the momentum space dual to the scattering region.

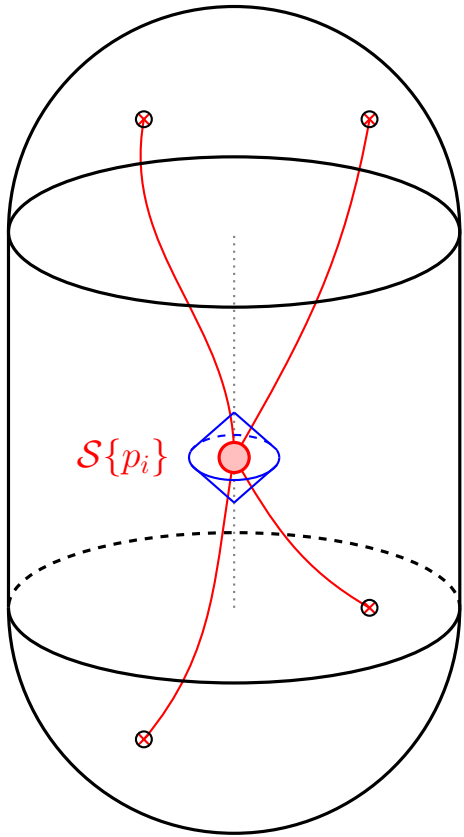
\begin{figure}[]
\centering

\begin{tikzpicture}
\draw[white,very thick,name path=TOP](1.25,7) arc (180:360:0.75 and 0.25);

\draw[gray, thick,dotted] (2,2) -- (2,7) node [pos=1,above=2]{{$ $}} node [pos=0,below=2]{{$ $}};

\draw[thick,red] (0.8,0.25) to[out=50,in=-100] (2,4);
\draw[thick,red] (3.8,2) to[out=150,in=-60] (2,4);
\draw[thick,red] (3.8,8.5) to[out=-100,in=60] (2,4);
\draw[thick,red] (0.8,8.5) to[out=-100,in=90] (2,4);

\draw[black,very thick,dashed](5,2) arc (0:180:3 and 1);
\draw[black,very thick](-1,2) arc (180:360:3 and 1);

\draw[blue,thick](2.6-0.03,4+0.1)--(2,4.6);
\draw[blue,thick](1.4+0.03,4+0.1)--(2,4.6);

\draw[blue,thick,dashed](2.6,4) arc (0:180:0.6 and 0.3);
\draw[blue,thick](2.6,4) arc (0:-195:0.6 and 0.3);

\draw[blue,thick](2.6-0.03,4-0.1)--(2,3.4);
\draw[blue,thick](1.4+0.03,4-0.1)--(2,3.4);

\draw[red,very thick,fill=pink] (2,4) circle (0.2) node[left=20]{\Large ${\cal S}\{p_i\}$};

\draw[black,very thick] (-1,2) -- (-1,7);
\draw[black,very thick] (5,2) -- (5,7) ;

\draw[black,very thick](2,7) ellipse (3 and 1);

\draw[black,very thick] (5,7) arc (0:180:3 and 3);
\draw[black,very thick] (5,2) arc (0:-180:3 and 3);

\draw[thick] (0.8,0.25) node[cross,red] {};
\draw[thick] (0.8,0.25) circle (3pt) node[right]{   };

\draw[thick] (3.8,2) node[cross,red] {};
\draw[thick] (3.8,2) circle (3pt) node[right]{   };

\draw[thick] (3.8,8.5) node[cross,red] {};
\draw[thick] (3.8,8.5) circle (3pt) node[right]{   };

\draw[thick] (0.8,8.5) node[cross,red] {};
\draw[thick] (0.8,8.5) circle (3pt) node[right]{   };

\end{tikzpicture}
    \caption{ Pictorial representation of the S-matrix according to formula \ref{eq:Smatrix}. We have chosen to represent a $2\rightarrow 2$ event where two ``in" particles of any mass are created in the scattering region and become two ``out" particles after interacting non-trivially in flat space. The interaction is represented by a red blob, and its details are encoded in the relevant CFT correlator in the form of a kinematic divergence.  The primary operators are located at insertions of the form \ref{eq:insertion}, whose complex value has been represented as an insertion in the Euclidean half-sphere. 
}\label{fig:SmatrixPic}
\end{figure}

After these manipulations, we can now simplify formula \ref{eq:Ppv1} and write the following correspondence between momentum operators in the flat space scattering region and smearings of CFT primaries at the boundary
\be\label{eq:Ppv2}
\hat{\phi}(p) \underset{l\rightarrow \infty}{=}e^{m l} \int dt  \, e^{\pm i \omega t} {\cal O}\left(   \tau   ,\Omega \right) \, .
\ee
with
\be\label{eq:insertion}
\tau =\pm{{\pi}\over 2} \pm i \cosh^{-1}{{\omega}\over{k}}   +{{t}\over l}   ,\quad \text{and}\quad  \Omega=\chi +{{\pi}\over 2}  \mp        {{\pi}\over 2} \, .
\ee
The overall exponential prefactor in \ref{eq:Ppv2} comes from the first term in equation \ref{eq:sigma}. Using formula \ref{eq:Ppv2} one can now write scattering amplitudes in the flat space geometry as follows; For a set of external particles with momenta $p_i$, the S-matrix associated to a physical process in flat space reads
\be\label{eq:Smatrix}
{\cal S}\{ p_i  \} =   \lim_{l\rightarrow\infty} l^{{d-3}\over 2} \left[ \prod_i   C(p_i) \int dt_i \, e^{\pm i \omega_i t_i}\right] \langle 0 \vert  {\cal O}\left(   \tau_1   ,\Omega_1 \right)   \cdots      \vert 0 \rangle\, .
\ee
Here, the primaries are located at insertions $\tau_i,\Omega_i$ of the form \ref{eq:insertion}, and the choice of sign in all these formulas represents the choice between ingoing and outgoing particles.  We have also introduced an overall power of the AdS length scale and defined the $p$-dependent factor 
\be\label{eq:normalization}
C(p_i)=e^{m_i l}\left(   {{2}\over {l k_i}}   \right)^{\Delta_i-1}\sqrt{  { \omega_i }\over{k_i}    }\, ,
\ee
which ensure that the resulting S-matrix is properly normalized. Formula \ref{eq:Smatrix} is the main result of this section. For a pictorial representation of this equation see figure \ref{fig:SmatrixPic}. The rest of the paper is devoted to testing this proposal for the simple correlators we will  derive in section \ref{sec:BMS3}, and for more complicated scattering events as shown in section \ref{sec:CONE}.

Before we proceed, some comments are in order. Firstly, formula  \ref{eq:Smatrix} has been obtained assuming the semi-classical limit of the HKLL formula without interactions. We would like to argue that this is indeed what one must do to build a scattering amplitude involving asymptotic Fock states. Generally, in order to build asymptotic states one must choose an asymptotic Hamiltonian that is simpler than the one governing finite-time dynamics. This is because next to the asymptotic boundary, particles are very far away from each other and interactions can be ignored. The choice of free Hamiltonian leads to the construction of a Fock space, and this is the case we consider throughout this work. As a consequence, we must construct asymptotic states using a free version of the HKLL formula that also ignores long range forces like gravity. It is worth mentioning that considering Fock space states leads to IR divergences. This might be solved by considering ``inclusive'' quantities \cite{Weinberg:1965nx}, or by considering ``dressed'' asymptotic states  \cite{Carney:2018ygh,Carney:2017oxp}. Such dressed states might be related to more complicated versions of the HKLL formula that include gravitational interactions. Alternatively, one might want to express local bulk operators using a different prescription (some 3D examples are cross-cap states \cite{Lewkowycz:2016ukf}, proto-fields \cite{Anand:2017dav,Chen:2019hdv}, or recovery channels \cite{Cotler:2017erl,Chen:2019gbt}). One could then obtain a formula similar to   \ref{eq:Smatrix}   but involving non-primary operators.  This is however beyond the scope of this paper. It is also important that the result presented in this section has been obtained for a pure AdS  background geometry. Formula \ref{eq:Smatrix} has to be slightly modified if the state of the CFT  is not the vacuum. This will be seen explicitly in the computations of section \ref{sec:CONE}, which involve a CFT$_2$  state dual to a conical deficit bulk geometry in AdS$_3$. 


\subsection{Two-point function}
In this subsection we test the proposal \ref{eq:Smatrix} for the case of two massive particles with momenta $p_1$ and $p_2$ in three dimensional Minkowski space ($d=2$)\footnote{
The computation in higher dimensions is conceptually equivalent. A spectral decomposition formula like \ref{eq:2ptSpectralvac} exists and involves higher dimensional spherical harmonics \cite{Higuchi:1986wu} that obey completeness relations leading to Dirac delta distributions like the ones leading to formula \ref{eq:S2ptLimit}.
}. The techniques used here will lay the groundwork for other computations in this paper. The S-matrix can be computed from the conformal two-point function through the formula
 \be
{\cal S}\{ p_1,p_2 \} = \lim_{l\rightarrow\infty}l^{-{1\over 2}}\left[\prod_{i=1}^2 C(p_i) \int dt_i\, e^{i \omega_i t_i}   \right] \langle 0 \vert  {\cal O}\left(   \tau_1   ,\phi_1 \right)    {\cal O}\left(   \tau_2   ,\phi_2 \right)        \vert 0 \rangle\, .
\ee
where we have chosen both particles to be outgoing, with
\be
\tau_i ={{\pi}\over 2} + i \cosh^{-1}{{\omega_i}\over{k_i}}   +{{t_i}\over l}   ,\quad \text{and}\quad  \phi_i=\chi_i  \, .
\ee
We will use the spectral form of the conformal two-point function
\be\label{eq:2ptSpectralvac}
 \langle 0 \vert  {\cal O}\left(   \tau_1   ,\phi_1 \right)    {\cal O}\left(   \tau_2   ,\phi_2 \right)        \vert 0 \rangle=\sum_{n\in \mathbb{Z}, \kappa\in \mathbb{Z}^+}  e^{i n \phi_{12}}e^{-i(\Delta+\vert n \vert +2\kappa)\tau_{12}} {{\Gamma(\vert n \vert +\Delta+\kappa)\Gamma(\kappa+\Delta)}\over{\Gamma(k+\vert n\vert+1)\kappa!}}\, .
\ee
The strategy is to perform the sum over $\kappa$ explicitly, which will result in a hypergeometric function. This function will then be written as an integral, which will be approximated through a stationary phase approximation in the large $l$ limit. The integrals over $t_i$ will then be performed, resulting in Dirac delta distributions that will come in handy to simplify the final answer. The sum over $\kappa$ yields
\be
 \langle 0 \vert  {\cal O}\left(   \tau_1   ,\phi_1 \right)    {\cal O}\left(   \tau_2   ,\phi_2 \right)        \vert 0 \rangle=\sum_{n\in \mathbb{Z}}  e^{i n \phi_{12}}  
z^{  \Delta+{{\vert n \vert}\over{\alpha}}   }    {{\Gamma\left(  \Delta  \right) \Gamma\left(  \Delta+ \vert n \vert  \right)    }\over{ \Gamma\left(  1+ \vert n \vert   \right) }} {}_2 F_1\left(   \Delta\, ,\,  \vert n \vert +\Delta\, ; \, \vert n \vert +1\, \vert \,   z^2     \right)
      \, .
\ee
where we have defined $z=e^{-i\tau_{12}}$.  The hypergeometric function allows for a complex integral representation.
\be
\langle 0 \vert  {\cal O}\left(   \tau_1   ,\phi_1 \right)    {\cal O}\left(   \tau_2   ,\phi_2 \right)        \vert 0 \rangle=\sum_{n\in \mathbb{Z}}  e^{i n \phi_{12}}  
z^{  \Delta+\vert n \vert  }  
 \int_{-i\infty}^{i \infty} {{ds}\over{2\pi i}}     {{\Gamma\left(   s  \right)   \Gamma\left(  \Delta-  s  \right)    \Gamma\left(  \vert n \vert+\Delta- s  \right)  }\over{  \Gamma\left(  \vert n \vert+1- s  \right)    }}    \left( -z^2      \right)^{-s}
\ee
The integral representation is valid as long as $\vert \text{arg}(-z^2) \vert<\pi$, which is true for our insertions as long as we do not take the strict $l\rightarrow \infty$ limit before performing the integral. We can now replace the values of $\tau_i$ and choose conformal dimensions $\Delta_i = m l$ such that the dual fields stay massive as we take the large $l$ limit. In such limit, the integrals over $t_{i}$ suggest that the integral over $s$ is dominated by values $s=l s'$. We can thus write
\be
{\cal S}\{ p_1,p_2 \} = \lim_{l\rightarrow\infty}l^{-{1\over 2}}\left[\prod_{i=1}^2 C(p_i) \int dt_i\, e^{i \omega_i t_i}   \right] \sum_{n\in \mathbb{Z}}  e^{i n \phi_{12}}   \int ds' e^{i t_-  \left(  2s'-m  \right)} g(s')    \,   e^{l f(s')}\, ,
\ee
where we have defined the following functions of $s'$
\be
\begin{split}
f(s')&=2 s' \log s' +2(m-s')\log(m-s') +(m-2s') \log {{k_1    \left(  \omega_2+\sqrt{\omega_2^2-k_2^2}    \right) }\over{k_2 \left(  \omega_1+\sqrt{\omega_1^2-k_1^2}    \right) }}\, , \\
g(s')&={1\over l} e^{-2 m l} l^{2m l} (-1)^{  -  \vert n \vert   }    (m-s')^{ \vert n \vert -1 } (s')^{-\vert n \vert-1}   \left(   {{k_1    \left(  \omega_2+\sqrt{\omega_2^2-k_2^2}    \right) }\over{k_2 \left(  \omega_1+\sqrt{\omega_1^2-k_1^2}    \right) }}    \right)^{  \vert n \vert }  \, .
\end{split}
\ee
The integral can then be evaluated using a stationary phase approximation. We look for saddle points obeying $\partial_{s'}f(s')=0$.  The saddle is located at
\be
s'_*=   {{m }\over{2(k_2\omega_1+k_1\omega_2)}}   \left[   k_2\left(  \omega_1-\sqrt{\omega_1^2-k_1^2}  \right)        + k_1\left(  \omega_2+\sqrt{\omega_2^2-k_2^2}  \right)         \right]\, .
\ee
And so the result for the integral over $s$ reads, as a function of $s'_*$, 
\be
{\cal S}\{ p_1,p_2 \} = \lim_{l\rightarrow\infty}l^{-{1\over 2}}\left[\prod_{i=1}^2 C(p_i) \int dt_i\, e^{i \omega_i t_i}   \right] \sum_{n\in \mathbb{Z}}  e^{i n \phi_{12}}   e^{i t_-  \left(  2s'_* -m  \right)} g(s'_*)   \sqrt{ {2\pi}\over{l f''(s'_*)}    } \,   e^{l f(s'_*)}\, .
\ee
This expression can be simplified greatly by performing the integrals over $t_i$, which are Dirac deltas enforcing conservation of the energy and radial momentum. After some simplification, we find
\be
{\cal S}\{ p_1,p_2 \} \sim \lim_{l\rightarrow\infty}\left[\prod_{i=1}^2 C(p_i) \right] \left[   e^{-m l} \left(   {{l k_1}\over 2}  \right)^{l m-1} \right]^2 {{k_1}\over{\omega_1}} \sum_{n\in \mathbb{Z}}  e^{i n \phi_{12}}  (-1)^{-\vert n \vert} \delta(\omega_1+\omega_2){1\over {k_1}}\delta(k_1-k_2) 
\ee
where we have ignored overall constants. Using now the expression for the normalizations $C(p_i)$ from equation \ref{eq:normalization} and performing the sum over $n$ yields simply
\be\label{eq:S2ptLimit}
{\cal S}\{ p_1,p_2 \} \sim\delta^{(3)}(p_1+p_2)\, . 
\ee
We conclude that non-zero scattering involves vanishing total momentum $p_1+p_2$, as we will also derive using Poincar\'e invariance in formula \ref{eq:S2ptLimit} of section \ref{sec:BMS3}. The result presented here is mathematically involved but conceptually trivial, and serves only as a consistency check for the proposal \ref{eq:Smatrix}, and also as a way to normalize the S-matrix. In the next subsection we analyze a more complicated correlator involving four CFT$_2$ operators.


\subsection{Four-point function: global BMS$_3$ blocks from conformal blocks }\label{sec:flatblocks}
We now focus on the S-matrix amplitude  involving four external particles. The proposal \ref{eq:Smatrix} involves the conformal four-point function. This correlator is not fixed by conformal invariance, and can be expanded in any basis of conformally invariant functions. Here, we will consider the basis involving conformal blocks. As we will show, considering the contribution from a particular block will result in an S-matrix resembling a BMS$_3$ block.  There objects will also be constructed from a BMS$_3$ theory perspective in section \ref{sec:BMS3}. Alternatively, instead of considering conformal blocks one can study conformal four-point functions obtained from holographic AdS$_3$ calculations like a Witten diagram.  This would result in S-matrix amplitudes  resembling Feynman diagrams. 

Some of the computations shown here have also been performed previously in the literature. In \cite{Penedones:2010ue,Fitzpatrick:2011hu,Fitzpatrick:2011dm}, S-matrices involving external massless particles were obtained from the conformal four-point function involving primaries with conformal dimensions $\Delta\sim{\cal O}(1)$. In \cite{Paulos:2016fap}, an expression for massive external particles ($\Delta=m l$) was proposed. In this note, it will be clear how both of these approaches arise from formula \ref{eq:Smatrix}. 

The Mellin representation of CFT correlators \cite{Mack:2009mi,Mack:2009gy} is central to the constructions presented in previous works, and we will also make use of it here.  The Mellin amplitude is an integral transform of a correlator in position space
\be\label{eq:mellin} 
\langle {\cal O}(x_1)...{\cal O}(x_n) \rangle = \int [d\gamma]\,  {\cal M}(\gamma_{ij}) \prod_{1\leq i\leq j\leq n}  \Gamma(\gamma_{ij}) \left(x_{ij}^2\right)^{-\gamma_{ij}}\, ,
\ee
where the Mellin variables obey $\gamma_{ii}=-\Delta_i$, $\gamma_{ij}=\gamma_{ji}$\, and $\sum_i \gamma_{ij}=0$, and the kinematic objects  in global coordinates are
\be
x_{ab}^2=2 \cos\tau_{ab}-2\cos\phi_{ab}\, .
\ee
These objects will have to be evaluated at the insertions \ref{eq:insertion} in order to calculate the S-matrix \ref{eq:Smatrix}. 

In this section, we will study the Mellin amplitude associated to a conformal block involving four operators exchanging a spinless representation with conformal dimension $\Delta'$. In the case of the four-point function, there are six Mellin variables, and the constraints fix four of them, leaving $\gamma_{24}$ and $\gamma_{34}$ unfixed. The explicit formulas are
\be
\begin{split}
\gamma_{12} &= \gamma_{34}+{{\Delta_1+\Delta_2-\Delta_3-\Delta_4}\over 2}\, ,\quad  \quad \quad \quad \gamma_{23}=\gamma_{24} +{{\Delta_1-\Delta_2+\Delta_3-\Delta_4}\over 2}\, , \\
\gamma_{23} &=-\gamma_{24}-\gamma_{34}+{{\Delta_2-\Delta_1+\Delta_3+\Delta_4}\over 2}\, \quad \gamma_{14}=-\gamma_{24}-\gamma_{34}+\Delta_4\, .
\end{split}
\ee
In terms of the remaining free Mellin variables, the amplitude reads \cite{Mack:2009mi,Fitzpatrick:2011hu}
\be\label{eq:MellinBlock}
{\cal M}(\gamma_{ij}) = C_{\Delta_i,\Delta'}e^{\pi i (h-\Delta')} \left( e^{i\pi (2s+ \Delta'-2h)}-1   \right)
{{ \Gamma\left( {{\Delta'}\over 2}  -s \right)\Gamma\left( {{2h-\Delta'}\over 2} -s  \right)}
\over
{ \Gamma\left( {{\Delta_1+\Delta_2}\over 2}-s  \right)\Gamma\left( {{\Delta_3+\Delta_4}\over 2}-s  \right)}}\, .
\ee
with $h=d/2=1$, 
\be
C_{\Delta_i,\Delta'}={{\Gamma\left(   \Delta_1 +\Delta_2-\Delta'   \right)\Gamma\left(   \Delta_1+\Delta_2+\Delta'   \right)\Gamma\left(   \Delta_3+\Delta_4-\Delta'   \right)\Gamma\left(     \Delta_3+\Delta_4+\Delta' \right)}\over{\Gamma\left(   \Delta_1  \right)\Gamma\left(  \Delta_2    \right)\Gamma\left(  \Delta_3    \right)\Gamma\left(   \Delta_4   \right)}}\, ,
\ee
and $s=-\gamma_{34}+(\Delta_3+\Delta_4)/2$.

The calculation of the S-matrix proceeds differently depending on whether we study massless or massive particles. We first study the massless case.

\subsubsection{Massless external particles}
In the case of particles with null momenta, we need to study primary operators dual to AdS bulk fields with zero mass in the large $l$ limit. We must have
\be
m^2=\lim_{l\rightarrow\infty}   {{\Delta(\Delta-d)}\over{l^2}} =0\, .
\ee
This is achieved by considering  operators with conformal dimension
\be
\Delta={\cal O}(1)\quad \text{as} \quad l\rightarrow\infty\, .
\ee
Without loss of generality, we will consider four primary operators with the same conformal dimension. For null momenta, we have $\omega=\mp k$. The insertions \ref{eq:insertion} in this case read
\be\label{eq:insertionMassless}
\tau_i =\pm{{\pi}\over 2}  +{{t_i}\over l}   ,\quad \text{and}\quad  \phi_i=\chi_i +{{\pi}\over 2}  \mp        {{\pi}\over 2} \, .
\ee
The imaginary part of global time vanishes and the primaries are smeared in a fringe of global time around $\tau=\pm \pi/2$. This is precisely the construction presented in \cite{Fitzpatrick:2011hu}\footnote{
In the calculation presented here we are focusing in (2+1)-dimensional Minkowski space. Nevertheless, formula \ref{eq:Smatrix} reduces to the construction presented in previous works in the literature in any space-time dimensions. For an explicit calculation of the S-matrix involving massless external particles in general dimensions, see section 2 in \cite{Fitzpatrick:2011hu}.
}. The insertions can be understood as creating a light particle in AdS that propagates along a null ray and hits the scattering region at the center of global AdS around global time $\tau\approx 0$. Inserting more operators of this kind in the CFT thus sets up a scattering event around the center of global AdS involving external massless particles. 

We have all the ingredients necessary to evaluate the proposal \ref{eq:Smatrix} for the scattering amplitude. We now use formula \ref{eq:mellin} for the CFT correlator, insert the primaries at the insertions \ref{eq:insertionMassless}, expand the integrand in the large AdS length scale limit, and perform the integrals over the time coordinates $t_i$. The details of this calculation can be found in previous works \cite{Fitzpatrick:2011hu}, and we have reviewed it in appendix \ref{app:massless}. Here we explicitly write the result. In terms of the Mellin amplitude, we obtain
\be
\begin{split}
{\cal S}\sim \, \delta^{3}\left(\sum_i p_i \right) 
\int d\alpha \, e^{\alpha} \alpha \, {\cal M}\left( \gamma_{24}=l^2 \delta_{24}^*, \gamma_{34}=l^2 {{s_{34}}\over{4\alpha}}\right)\, .
\end{split}
\ee
Using now the explicit expression for the Mellin amplitude of a conformal block and approximating the result using Stirling's formula yields
\be\label{eq:MasslessFinal}
{\cal S}\sim \delta^{(3)}\left(   \sum_{i} p_i   \right) \delta\left(  s+M^{\prime 2}    \right)\, ,
\ee
which is precisely the formula for the global BMS$_3$ block. This formula will also be computed in  \ref{eq:block}  from the perspective of a theory with BMS$_3$ symmetry.

\subsubsection{Massive external particles}
We turn our attention to external particles with non-zero mass. In this case, a finite mass for the AdS bulk field can be achieved by studying primaries  with conformal dimension
\be
\Delta_i = m_i l\, .
\ee
The primaries have to be inserted at the locations \ref{eq:insertion} for time-like momenta $p_i$. In this case the insertion is complex in global time. This makes sense by recalling that massive fields propagating in anti-de Sitter space-time cannot reach the boundary of space with finite energy. As drawn in figure \ref{fig:SmatrixMap}, Inserting the CFT operators in the Euclidean half sphere creates massive fields that propagate through AdS and focus on a narrow wave-packet when approaching the flat space scattering region. 

We have all ingredients necessary to write down the integrand in our formula \ref{eq:Smatrix} and perform the integrals over the Mellin variables and the times $t_i$. The computation is conceptually simple but technically involved, so we have relegated a summary of the main steps in appendix \ref{app:massive}. The result involves a saddle point approximation to the Mellin integrals, whose saddle reads
\be\label{eq:saddlemain}
\gamma_{ab}^*=l {{m_a m_b+p_a\cdot p_b}\over{\sum_i m_i}}\, .
\ee
This is precisely the point in Mellin space relevant for the proposal in \cite{Paulos:2016fap}. We conclude that the formula \ref{eq:Smatrix} roughly reduces to the proposal in \cite{Paulos:2016fap} in the case of massive external particles. Evaluating the Mellin integrals at the saddle \ref{eq:saddlemain}, the Dirac integrals over $t_i$, and performing some algebra, one can conclude that for the block Mellin amplitude  \ref{eq:MellinBlock}, the resulting scattering amplitude reads
\be
{\cal S}\sim \delta^{(3)}\left(   \sum_{i} p_i   \right) \delta\left(  s+M^{\prime 2}    \right)\, ,
\ee
which matches again the formula for a global BMS$_3$ block (see section \ref{sec:BMS3}). Note that this formula is slightly different from the one found for the massless case, as the momenta appearing here are time-like on-shell vectors obeying $p^2_i=-m_i^2$.

The results presented in this section are the simplest applications of the S-matrix proposal. In the next section, we will interpret the results from the point of view of a BMS$_3$ theory that could represent the holographic dual to asymptotically flat space-times in $(2+1)$ dimensions. A slightly more complicated application of the proposal can be found in section \ref{sec:CONE}, which studies flat limits of CFT$_2$ correlators in states different from the vacuum.


\section{Asymptotically flat space-times and BMS$_3$ symmetry}\label{sec:BMS3}
In this section we analyze the asymptotic structure of flat space-time in three dimensions. The objective is to review the construction of the asymptotic symmetry group and its unitary representations, as well as the introduction of bms$_3$ operators that will lead to the definition of BMS correlators. The simplest correlators fixed by global symmetry will then be compared with the flat limit results obtained in the previous section.

We consider Einstein gravity with vanishing cosmological constant in $(2+1)$ dimensions. In Eddington-Fikelstein coordinates, the line element for the vacuum solution reads
\be
ds^2=-du^2-2du dr +r^2d\phi^2\, .
\ee
where $u=t-r$ is the retarded time, and $\phi\sim\phi+2\pi$. The future null boundary is located at $r\rightarrow\infty$, with $u,\phi$ fixed. The charges associated to the asymptotic killing vectors of this geometry obey a centrally extended algebra associated to the BMS$_3$ group. The commutators read
\be\label{eq:algebra}
\begin{split}
[L_n,L_m]&=(n-m)L_{n+m} +\mathbf{1}{{c_L}\over {12}}n(n^2-1)\delta_{n+m}\, , \\
[L_n,M_m]&=(n-m)M_{n+m} +\mathbf{1}{{c_M}\over {12}}n(n^2-1)\delta_{n+m}\, , \\
[M_n,M_m]&=0\, .\\
\end{split}
\ee
Here $L_n$ stand for generators of super-rotations, while $M_n$ correspond to generators of super-translations. The central charges depend on the specific theory of gravity. For the case of Einstein gravity,
in order for the phase space of three-dimensional asymptotically flat gravity to match the
space of coadjoint representations of the BMS$_3$ group, it is required that \cite{Barnich:2014kra}
\be
c_L=0\, , \quad \text{and}\quad c_M={3\over {G_N}}\, .
\ee
The application of the holographic principle in this context implies the existence of a theory whose physical content organizes into irreducible representations of the algebra \ref{eq:algebra}. We proceed to discuss such representations.

\subsection{Unitary representations of BMS$_3$}\label{sec:UniRep}
In this paper we will study the unitary representations of the BMS$_3$ group, which should be the relevant representations when discussing holography. For a careful discussion concerning many of the details discussed in this subsection, see \cite{Campoleoni:2016vsh} and \cite{Barnich:2014kra}. A class of  representations that we consider in this work are provided by the Hilbert space defined through the states $\vert M, s\rangle$ obeying
\be\label{eq:StateMs}
M_0\vert M, s\rangle = M \vert M, s\rangle\, , \quad M_{n\neq 0}\vert M, s\rangle=0\, , \quad \text{and} \quad L_0 \vert M, s\rangle = s \vert M, s\rangle\, .
\ee
The states along the representation labeled by the quantum numbers $M$, $s$ are obtained by acting with generators of super-rotations $L_{n\neq 0}$. The representations considered here can be seen to arise from an ultra-relativistic limit of the highest weight representations of the symmetry algebra of a two-dimensional conformal field theory, which consists on two copies of the Virasoro algebra. If the Virasoro algebras are spanned by generators ${\cal L}_n$, $\bar{\cal L}_n$, and central charges $c$ and $\bar{c}$, the contraction reads 
\be
M_n = {1\over l}\left(  {\cal L}_n+\bar{\cal L}_{-n}  \right)\, , \quad L_n = {\cal L}_n -\bar{\cal L}_{-n}\, ,
\ee
and 
\be
c_M=c+\bar{c}\, , \quad c_L= c-\bar{c}\, ,
\ee
where $l$ can be thought of as the AdS length scale, which is taken to infinity. Under this contraction, the highest weight representations with conformal dimensions $h,\bar{h}$ map to unitary massive modules with the following quantum numbers
\be
M={{h+\bar{h}}\over l}\, , \quad s=h-\bar{h}\, .
\ee
The existence of a limit relating unitary representations of the Virasoro algebra to unitary representations of the BMS$_3$ algebra can be regarded as evidence that the observables in the CFT$_2$ can be related to observables in a theory with BMS$_3$ symmetry.  Before addressing this issue, we first need to define operators in a theory with BMS$_3$ symmetry. 


\subsection{BMS$_3$ operators}\label{sec:StateOp}
In the context of AdS/CFT, a set of observables commonly studied are conformal correlators \cite{Harlow:2011ke}. In order to compute these objects, one must first define the operators appearing in the correlator. This is done through the state-operator map, relating states in a given highest weight representation to a local insertion of an operator in the manifold where the CFT is defined. In the case of a theory with BMS$_3$ symmetry, we have defined the relevant representations in section \ref{sec:UniRep}. We are now ready to study the nature of the operators associated to states in such representations. 

States in unitary representations can be understood as insertions of local operators in super-momentum space. The generators $L_n$, $M_n$, and $\mathbf{1}$ appearing in the algebra \ref{eq:algebra} naturally induce the adjoint representation of the bms$_3$ algebra acting on the linear operators of the Hilbert space associated to a massive module. The adjoint representation acts as the commutator. Denoting the generators of the symmetry as $G_a$, and a linear operator as $\Phi$, we have
\be
\text{Ad}_{G_a}\Phi = [G_a,\Phi]\, .
\ee
This directly implies that the center generators have vanishing action on the operator $\Phi$, so that the adjoint representation of the generators of the algebra is a representation of the non-centrally extended bms$_3$ algebra, which we will denote $\hat{\text{bms}}_3$.

The $\hat{\text{bms}}_3$ algebra admits different geometric realizations. In some works \cite{Bagchi:2017cpu,Hijano:2017eii,Hijano:2018nhq}, a realization involving differential operators in the plane has been used to construct local operators that are associated to states in highest weight representations. For the modules specified in section \ref{sec:UniRep}, the natural realization of the algebra is in super-momentum space. The algebra consists of differential operators acting on functions of super-momentum coordinates $p^{(n)}$. The realization is the following. 
\be\label{eq:Preal}
\begin{split}
m_{n}&= p^{(n)}\, , \quad \text{and} \quad l_n= \sum_{m\in\mathbb{Z}}  (m-n) p^{(m+n)} \partial_{p^{(m)}}\, .
\end{split}
\ee
The global generators of translations are realized by $p^{(n)}$ with $n=\pm 1,0$, which are coordinates in momentum space. Explicitly,
\be
p^{(0)}=\omega\, \quad \text{and} \quad p^{(\pm 1)}=k e^{\pm i \chi}\, ,
\ee
where the Lorentz vector $(\omega,\mathbf{k})$ stands for thee-momentum, $k=\vert \mathbf{k}\vert$, and $\chi$ is an element of the unit circle.

In order to define an operator at a point in momentum space, we first need to find a point stabilized by the generators $l_0$ and $m_n$, which are the generators needed to define the representation algebraically in \ref{eq:StateMs}. This point is given by $p^{(n\neq 0)}=0$. Demanding that the action of $M_0$ on the operator placed at this point is consistent with the algebraic definition of the state $\vert M,s\rangle$ in \ref{eq:StateMs} yields $p^{(0)}=M$. It is clear then that the state $\vert M,s\rangle$ corresponds to the insertion of an operator at the rest-frame (super-)momentum $p_0^{(n)}=M\delta_{n,0}$.  We thus state the following correspondence
\be
\Phi_s(p_0) \vert 0 \rangle \equiv \vert M,s\rangle\, ,
\ee
where, $\vert 0 \rangle$ stands for the Poincar\'e  invariant vacuum, which corresponds to the $M=s=0$ choice of quantum numbers in the massive module.

Having found a point in super-momentum space in which to insert an operator, we now need to act with global super-rotations to move the operator to a general point in momentum space. The logic is equivalent to the construction of local operators in conformal field theories, where operators are first defined at the origin, and then moved away from the origin by using global conformal transformations. Operators\footnote{Note that when super-momentum is turned off, the operators $\Phi(p)$ discussed here are equivalent to the creation operators $a^{\dagger}_p$ appearing in the Wigner representation of massive particles in three dimensions.} at a general point in momentum space are then
\be
\Phi(p) = U^{-1} \Phi(p_0) U\, , \quad \text{with} \quad U=\text{Exp}\left(   \sum_{n=-1}^{n=1} \omega_n  L_n  \right)\, ,
\ee
where the parameters $\omega_n$ parametrize the Fourier expansion of a global super-rotation vector field on the circle $\omega(\phi)\partial_{\phi}$. Knowing the action of the generators of the symmetry on these fields at the rest frame now allows us to derive how the operators transform  in general points in momentum space. Explicitly, we obtain
\be\label{eq:TransfP}
\begin{split}
[M_{n},\Phi(p) ]&= p^{(n)}  \Phi(p) \quad \text{for}\quad n=-1,0,1\, , \\
[L_n, \Phi(p)] &= \sum_{m=-1}^{m=1}  (m-n) p^{(m+n)} \partial_{p^{(m)}} \Phi(p) \quad \text{for}\quad n=-1,0,1\, .
\end{split}
\ee
More general insertions involving non-trival super-momentum correspond to descendant operators and descendant states of $\vert M,s\rangle$.  This is discussed very briefly in appendix \ref{app:desc}. Having discussed bms$_3$ operators, we are now ready to compute simple correlators.


\subsection{BMS$_3$ correlators}
Having defined BMS$_3$ operators in the last section, we can now assemble correlators of the form
\be
\langle 0 \vert \Phi(p_1)\Phi(p_2) \cdots \vert 0 \rangle\, .
\ee
These objects are heavily constrained by the symmetry of the theory, much like the case of conformal correlators in a CFT. Correlators involving two or three bms$_3$ operators are indeed completely fixed by the constraint of invariance under the global part of the symmetry. Acting with the global generators on the two-point function and using the transformation laws \ref{eq:TransfP} yields simple differential equations that can be solved to obtain
\be\label{eq:2pt}
\langle 0 \vert \Phi(p_1)\Phi(p_2)   \vert 0 \rangle \sim \delta^{(3)}(p_1+p_2)\, .
\ee
A similar treatment of the three-point function yields 
\be\label{eq:3pt}
\langle 0 \vert \Phi(p_1)\Phi(p_2)  \Phi(p_3)  \vert 0 \rangle \sim \delta^{(3)}(p_1+p_2+p_3)\, .
\ee
The first correlator not completely fixed by global BMS$_3$ invariance is the four-point function. Invariance under translations implies conservation of momentum
\be
\sum_{i=1}^{4} p^{(n)}_i =0 \, , \quad \text{for} \quad n=-1,0,1\, .
\ee
Invariance under Lorentz transformations implies that the correlator can only depend on two Mandelstam variables associated to the momenta $p_i$. Even though the correlator cannot be completely fixed, we can expand it in a basis of BMS$_3$ invariant functions that we will denote as BMS$_3$ blocks. The logic is exactly the same as the one leading to Virasoro blocks in the context of conformal field theories. Under a choice of channel, we insert the identity operator as a sum over a complete set of states
\be
\langle 0 \vert \Phi(p_1)\Phi(p_2)  \Phi(p_3)   \Phi(p_3) \vert 0 \rangle=\sum_{\alpha} \langle 0 \vert \Phi(p_1)\Phi(p_2)   \vert\alpha\rangle\langle\alpha\vert  \Phi(p_3)   \Phi(p_3) \vert 0 \rangle \equiv \sum_{\alpha} {\cal F}_{\alpha}\, .
\ee
The index $\alpha$ labels all representations associated to the physical content of the theory, and we have denoted the contribution from a given representation by ${\cal F}_{\alpha}$. Some of those states might corresponds to states in a massive module of the form considered in section \ref{sec:UniRep}. All states in a given massive module share the same eigenvalue with the casimir operators
\be
{\cal C}_1=M_0^2-M_{-1}M_1 \, , \quad \text{and} \quad {\cal C}_2=2L_0M_0-{1\over 2}\left(   L_{-1}M_1+L_1M_{-1}+M_1 L_{-1}+M_{-1}L_1   \right)\, .
\ee
In particular, the eigenvalue of the casimir ${\cal C}_1$ is the mass squared $M_{\alpha}^2$, while the eigenvalue with ${\cal C}_2$ stands for the helicity. Demanding that the function ${\cal F}_{\alpha}$ is an eigenvalue of the two casimirs completely fixes its dependence on the two independent Maldestam variables. If we consider four spinless fields exchanging a spinless representation, the result reads
\be\label{eq:block}
{\cal F}_{\alpha} \sim \delta^{(3)}\left(   \sum_{i} p_i   \right) \delta\left(  s+M_{\alpha}^2    \right)\, .
\ee
The results \ref{eq:block}, \ref{eq:3pt}, and \ref{eq:2pt} look very familiar in the context of quantum field theory in flat space. They correspond to the S-matrix amplitudes of particles propagating in a flat space geometry\footnote{Note that the expressions we have defined in this section are just a basis of BMS$_3$ invariant functions in which one can expand general correlation functions. More familiar amplitudes obtained by the evaluation of Feynman diagrams can be thought as a different basis, and they can be expanded into a linear combination of the blocks presented throughout this note}. Indeed, we can regard the correspondence between a gravitational theory with flat asymptotics and a theory with BMS$_3$ symmetry as a matching between BMS$_3$ correlators as defined in this section, and scattering amplitudes in an asymptotically flat geometry. Unfortunately, as of now no theory with BMS$_3$ symmetry and a large central charge is known that could realize the holographic principle for physics in flat space in the limit of small Newton's constant. The purpose of this paper is to instead provide with an alternative microscopic understanding of the gravitational theory, as a ``flat'' limit of a conformal field theory. As we have seen in section \ref{sec:SMATRIX}, flat limits of global conformal correlators turn into the results  \ref{eq:block}, \ref{eq:3pt}, and \ref{eq:2pt} presented in this section. In the remainder of this paper, we study slightly more complicated CFT$_2$ correlators that result into non-trivial scattering amplitudes in flat space-time.


\section{Scattering on a cone}\label{sec:CONE}
In this section we seek to show how studying flat limits of non-trivial CFT$_2$ correlators can give rise to non-trivial scattering amplitudes in an asymptotically flat space-time. More specifically, we seek to obtain the S-matrix amplitude of a particle scattering against a cone geometry from an indirect, holographic perspective. In the conformal field theory, we will consider correlators involving primary operators in a state dual to a conical deficit AdS$_3$ geometry. 

The section is split as follows. We will first analyze the problem from the point of view of a quantum field theory living on an asymptotically flat geometry. We will then turn to the CFT$_2$ persepective and analyze correlators in deficit states. Finally, We will obtain the same answers by using a slightly modified version of formula \ref{eq:Smatrix}.


\subsection{Scattering against a cone in flat space-time}\label{sec:coneflat}
\begin{figure}[]
\centering
\begin{tabular}{ccc}
\begin{subfigure}[t]{0.31\textwidth}
\centering
\begin{tikzpicture}[scale=2.7]
\draw [white, thick] (-1,-1)--(1,1);
\draw [white, thick] (-1,1)--(1,-1);

\draw [blue, thick,dashed,-dot2-=0] (0,0)--(-1,0);

\draw [red, very thick,->] (0.5,-0.75)   to[out=90,in=-20]  (-0.25,0.75);
\end{tikzpicture}
\caption{}
\end{subfigure}\hspace{2mm}
&
\begin{subfigure}[t]{0.31\textwidth}
\centering
\begin{tikzpicture}[scale=2.7]
\draw [white, thick] (-1,-1)--(1,1);
\draw [white, thick] (-1,1)--(1,-1);

\draw [blue, thick,dashed,-dot2-=0] (0,0)--(-1,0);

\draw [red,thick,>=latex,->] (0,-0.5)--(0,-0.45);
\draw [red,thick,snake it] (0,-1)--(0,-0.5);

\draw [red,thick,>=latex,->] (-0.5,-0.5)--(-0.5,-0.45);
\draw [red,thick,snake it] (-0.5,-1)--(-0.5,-0.5);

\draw [red,thick,>=latex,->] (0.5,-0.5)--(0.5,-0.45);
\draw [red,thick,snake it] (0.5,-1)--(0.5,-0.5);
\end{tikzpicture}
\caption{}
\end{subfigure}
&
\begin{subfigure}[t]{0.31\textwidth}
\centering
\begin{tikzpicture}[scale=2.7]
\draw [white, thick] (-1,-1)--(1,1);
\draw [white, thick] (-1,1)--(1,-1);

\draw [blue, thick,dashed,-dot2-=0] (0,0)--(-1,0);

\draw [red,thick,>=latex,->] (0,1.5-0.5)--(0,1.5-0.45);
\draw [red,thick,snake it] (0,1.5-1)--(0,1.5-0.5);

\draw [red,thick,>=latex,->] (-0.5,1.5-0.5)--(-0.5,1.5-0.45);
\draw [red,thick,snake it] (-0.5,1.5-1)--(-0.5,1.5-0.5);

\draw [red,thick,>=latex,->] (0.5,1.5-0.5)--(0.5,1.5-0.45);
\draw [red,thick,snake it] (0.5,1.5-1)--(0.5,1.5-0.5);

\draw[green, very thick, ->] (0.75,0)--(0.8,0);
\draw[green, very thick, snake it] (0.25,0)--(0.75,0);

\draw[green, very thick, ->] (-0.75,0)--(-0.8,0);
\draw[green, very thick, snake it] (-0.25,0)--(-0.75,0);

\draw[green, very thick, ->] (0.375,0.650)--(0.4,0.693);
\draw[green, very thick, snake it] (0.125,0.216)--(0.375,0.650);

\draw[green, very thick, ->] (-0.375,0.650)--(-0.4,0.693);
\draw[green, very thick, snake it] (-0.125,0.216)--(-0.375,0.650);

\draw[green, very thick, ->] (0.375,-0.650)--(0.4,-0.693);
\draw[green, very thick, snake it] (0.125,-0.216)--(0.375,-0.650);

\draw[green, very thick, ->] (-0.375,-0.650)--(-0.4,-0.693);
\draw[green, very thick, snake it] (-0.125,-0.216)--(-0.375,-0.650);

\draw[darkgreen,opacity=0.5] (0,0) circle (0.30);
\draw[darkgreen,opacity=0.5] (0,0) circle (0.70);
\end{tikzpicture}
\caption{}
\end{subfigure}
\end{tabular}
    \caption{Free particles/fields ({\color{red} red}) scattering around a cone geometry ({\color{blue} blue}). The cone is represented in blue, with a dashed blue branch-cut representing the lack of periodicity under $\phi\sim\phi+2\pi$. \textbf{a)} Deflection of the trajectory of a classical particle that propagates freely in the conical deficit geometry.  \textbf{b)} Quantum wave-packets approach the cone source at $t=-\infty$, initially unaffected by the presence of the conical deficit. \textbf{c)}   As the waves interact with the source of the cone, they scatter. Parts remains a plane-wave ({\color{red} red}) that ignore the non-trivial geometry, while some of the wave scatters spherically ({\color{darkgreen} green}) away from the location of the source. 
}\label{fig:conepic}
\end{figure}
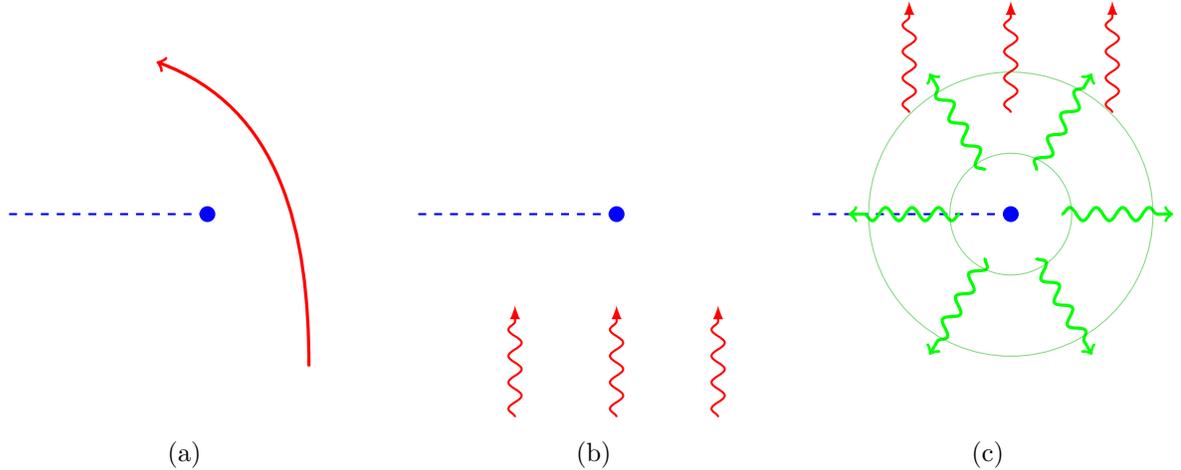

The problem of scattering in cone geometries was first studied non-relativistically in \cite{Deser:1988qn,tHooft:1988qqn}. The relativistic version was studied more recently in \cite{Spinally:2000ii}, and we will revisit the problem here in a different language. A general asymptotically flat geometry in 2+1 dimensions reads
\be\label{eq:Aflat}
ds^2=\Theta(\phi) du^2-2dudr+2\left[ \Xi(\phi)+{u\over 2}\partial\Theta(\phi)    \right]du d\phi+r^2 d\phi^2\, .
\ee
The choice $\Xi(\phi)=0$ and $\Theta(\phi)=-\alpha^2$ corresponds to an idealized cosmic string sitting at $r=0$. The line element is
\be\label{eq:ds2CS}
ds^2=-\alpha^2 du^2-2dudr+r^2 d\phi^2\, .
\ee
Fields propagating in this geometry effectively sense a conical deficit sourced at the origin. This can be seen explicitly by changing variables as
\be
 u\rightarrow {{t-r}\over {\alpha}} \,  \quad \text{and}\quad r\rightarrow\alpha r \, ,
\ee
which results in 
\be\label{eq:ds2cone}
ds^2=-dt^2 +dr^2 +r^2\alpha^2 d\phi^2\, .
\ee
As the angle coordinate is identified as $\phi\sim\phi+2\pi$, this geometry shows a conical deficit $\Delta \phi = \pi(\alpha^{-1}-1)$. A particle propagating in this geometry classically will follow a trajectory that is deflected by $\Delta \phi $, as shown pictorically in figure \ref{fig:conepic}.a).

We would like to understand the scattering quantum mechanically and relativistically, by studying quantum field theory in the background geometry \ref{eq:ds2cone}. We will compute the S-matrix amplitude by applying the LSZ reduction formula to a momentum space propagator in the cone geometry. The propagator of a scalar field of mass $m$ \cite{Moreira:1995he,Arefeva:2016wek} must obey 
\be\label{eq:DalembertianG}
\left(\nabla^2+m^2  \right) G(x,x')={1\over{\sqrt{-g}}} \delta^{(3)}(x-x')\, ,
\ee
and single-valuedness around the angle coordinate implies
\be\label{eq:SVG}
G(\phi+2\pi)=G(\phi)\, .
\ee
Solving \ref{eq:DalembertianG} with the single-valuedness condition \ref{eq:SVG} is a lengthy computation that can be found in appendix \ref{app:Gcone}. The resulting propagator captures all the details concerning the scattering of a free scalar field on an asymptotically flat conical defect background. However, here we are concerned with S-matrix elements that we can compare to correlators of BMS$_3$ operators. In Minkowski space, S-matrices are obtained by applying the LSZ reduction formula to momentum space correlators, which are obtained from position space correlators by Fourier transforming against plane-waves. When studying QFT in curved space, one cannot define plane-waves, as these cannot be constructed from eigenfunctions of the D'Alembertian \cite{Hollands:2014eia}. This is however not an issue here. S-matrix amplitudes are objects constructed at the null asymptotic boundary of Minkowski space at $r\rightarrow \infty$. The class of geometries \ref{eq:Aflat} are precisely defined such that the structure of the geometry at null infinity is that of Minkowski space. More concretely, the geometry \ref{eq:ds2cone} representing a cosmic string can be thought as the solution to the Einstein field equations in presence of a source at the origin $r=0$ \cite{Vilenkin:1984ib,Gott:1984ef}, such that local fields infinitely far away from the origin do not feel the presence of the string. All in all, this means that scattering amplitudes can be defined from propagators in asymptotically flat geometries by Fourier transforming against the plane-waves of Minkowski space.  We thus have
\be
G(p,p')=\int d^3 x \, e^{i \left(  -\omega t + k r \cos(\phi-\chi)  \right)}  \int d^3 x' \, e^{i \left(  -\omega' t' + k' r' \cos(\phi'-\chi')  \right)} \, G(x,x')\, .
\ee
The integrals over space-time points $x$ and $x'$ are performed explicitly in appendix \ref{app:Gcone}. The result can be written as follows
\be\label{eq:coneresult}
\begin{split}
G(p,p')&=\delta^{(3)}(p+p')+\delta(\omega+\omega'){1\over{k}}\delta(k-k')f(\chi,\chi')
\end{split}
\ee
where the momentum vectors appearing in the expression are on-shell, obeying $p^2=-m^2$, and the function $f(\chi,\chi')$ reads
\be
f(\chi,\chi')=  i {{\sin{{\pi}\over {\alpha}}}\over{  \cos{{\pi}\over{\alpha}}    - \cos(\chi-\chi')     }}\, .
\ee
 The interpretation is simple. As a particle propagates towards the cone as a plane-wave, the scattering event results in an outgoing plane-wave that must conserve momentum, and an outgoing spherical wave that captures the interaction with the conical defect. These two waves correspond to the two terms appearing in formula \ref{eq:coneresult}. As can be seen explicitly, turning off the cone by setting $\alpha=1$ results on a vanishing scattered spherical wave. A pictorial representation of this process has been drawn in figure \ref{fig:conepic}. 


\subsection{Correlator in a CFT$_2$ deficit state}
In this section we discuss correlators of operators in CFT$_2$ states dual to AdS$_3$ conical deficit geometries. Such states can be understood, in the large central charge limit, as created by the insertion of a heavy operator whose conformal dimension scales with the central charge. Naively, one could claim that the correlator of two light primaries is computed by the Virasoro vacuum block between the two light primaries and the two heavy primaries. This is however incorrect, and can be argued simply by the fact that Virasoro vacuum blocks are not single-valued as we move the light primaries around the insertion of the heavy operators \cite{Hijano:2015qja}. However, the holographic calculation  must involve a single-valued propagator, much like the flat space calculation shown in appendix \ref{app:Gcone}. The issue can be resolved by considering vacuum blocks in all channels, and summing over all channels in order to obtain a single-valued answer. This is the same logic presented in \cite{Maloney:2016kee} when building crossing symmetric correlators. For the case at hand, we claim
\be
\langle \alpha \vert {\cal O}(\tau,\phi) {\cal O}(\tau',\phi') \vert \alpha \rangle = \sum_{n=-\infty}^{\infty} {\cal V}_{\alpha} \left(  \tau-\tau',\phi-\phi'+2\pi n   \right)\, ,
\ee
where ${\cal V}_{\alpha}$ is the Virasoro vacuum block in the direct channel exchanging the identity representation between the light operators and the heavy operators associated to the deficit state $\vert \alpha \rangle$. Using the explicit expression for the vacuum block, we have
\be\label{eq:2ptConeCFTv0}
\langle \alpha \vert {\cal O}(\tau,\phi) {\cal O}(\tau',\phi') \vert \alpha \rangle= \sum_{n} \left({1\over{2\cos[\alpha(\tau-\tau')]-2\cos[\alpha(\phi-\phi'+2\pi n)]}}\right)^{\Delta} \, .
\ee
This result can be seen to be correct by solving the holographic correlator. The calculation can be found in appendix \ref{app:AdScone}. An alternative useful formula can be achieved by rewriting \ref{eq:2ptConeCFTv0} in a spectral decomposition
\be\label{eq:2ptConeCFT}
\begin{split}
\langle \alpha \vert {\cal O}(\tau,\phi) {\cal O}(\tau',\phi') \vert \alpha \rangle&= \sum_n\sum_{\kappa=0}^{\infty}    e^{i n (\phi-\phi')}  e^{-i\alpha\left( \Delta+{{\vert n \vert}\over{\alpha}}+ 2\kappa  \right) (\tau-\tau')}  
 {{\Gamma\left({{|n|}\over{\alpha}}+\Delta+\kappa\right)\Gamma\left(\Delta+\kappa\right)}\over{\Gamma\left({{|n|}\over{\alpha}}+\kappa+1\right)  \kappa ! }} \, .
\end{split}
\ee
This is the formula that will be used in the calculations of this section.

\subsection{Scattering on a cone as a flat limit of a CFT correlator}
In this section we aim to test the proposal \ref{eq:Smatrix} in the case of correlators in CFT states different from the vacuum. In appendix \ref{app:Gcone} we have revisited the problem concerning light particles scattering over asymptotically flat conical geometries. Our objective now is to re-derive the same result by taking a flat limit of a conformal correlator in a CFT deficit state generated by the insertion of a heavy operator.  Our proposal for the scattering amplitude reads
\be
{\cal S}\{ p_1,p_2 \} =l^{-{1\over 2}}\left[ \prod_{i=1}^2 C(p_i) \int dt_i \, e^{\pm i \omega_i t_i}\right] \langle \alpha \vert  {\cal O}\left(   \tau_1   ,\phi_1 \right)    {\cal O}\left(   \tau_2   ,\phi_2 \right)       \vert \alpha \rangle\, .
\ee
where $C(p_i)$ where defined in equation \ref{eq:normalization}, and the operators in the deficit correlator are inserted at
\be
\alpha\tau_i ={{\pi}\over 2} + i \cosh^{-1}{{\omega_i}\over{k_i}}   +{{t_i}\over l}   ,\quad \text{and}\quad  \phi_i=\chi_i  \, .
\ee
Note that this insertion slightly differs from the proposed insertions \ref{eq:insertion}.  The reason is that the AdS conical deficit geometry studied in formula \ref{eq:AdSConeGeom} results in the conical defect geometry studied in formula \ref{eq:ds2cone} upon taking the flat limit
\be
\alpha \tau =  {t\over l}\, , \quad \text{and} \quad \rho={r\over l}\, ,
\ee
which differs from the flat limit taken in formulas \ref{eq:FlatRegion} by the factor of $\alpha$. Otherwise the flat limit we propose remains unchanged. The correlator appearing here has been computed in the previous subsection so we have all ingredients needed to perform the calculation, which reads 
\be
{\cal S}\{ p_1,p_2 \} =\lim_{l\rightarrow\infty}l^{-{1\over 2}}\left[ \prod_{i=1}^2 C(p_i) \int dt_i \, e^{ i \omega_i t_i}\right] 
\sum_n\sum_{\kappa=0}^{\infty}    e^{i n (\chi_1-\chi_2)}  e^{- i  \left( \Delta+{{\vert n \vert}\over{\alpha}}+ 2\kappa  \right) \alpha \tau_{12}} 
C_{\kappa,n}
\, ,
\ee
where we have defined
\be
\begin{split}
C_{n,\kappa}&=   
 {{\Gamma\left({{|n|}\over{\alpha}}+\Delta+\kappa\right)\Gamma\left(\Delta+\kappa\right)}\over{\Gamma\left({{|n|}\over{\alpha}}+\kappa+1\right)  \kappa ! }}    \, .
\end{split}
\ee
The conformal dimensions of the operators in the correlator are taken to be $\Delta_i=m_i l$, such that the resulting particles are massive like the ones studied in section \ref{sec:coneflat} above. The strategy to perform this calculation is to re-write the sum over $\kappa$ as an integral over Mellin-like variables, resembling the set-up of section \ref{sec:flatblocks}. The resulting integral can then be performed in the large $l$ limit as a saddle point approximation. All the details have been relegated to appendix \ref{app:flatlimitcone}, while here we quote the final result
\be
{\cal S}\{ p_1,p_2 \} \sim \delta^{(3)}(p_1+p_2)+\delta(\omega_1+\omega_2){1\over k_1}\delta(k_1-k_2)  
{{i\sin{{\pi}\over{\alpha}}}\over{\cos{{\pi}\over{\alpha}}-\cos\chi_{12}}}\, .
\ee
This scattering matrix is precisely the one found in section \ref{sec:coneflat} and more explicitly in appendix \ref{app:Gcone} when studying light particles propagating in an asymptotically flat geometry. We can regard this result as the statement that the proposal \ref{eq:Smatrix} can relate non-trivial CFT$_2$ correlators to non-trivial scattering events in asymptotically flat geometries. We can thus analyze physics in asymptotically flat space-times through a holographic lens by studying observables in conformal field theories and transforming them into S-matrix amplitudes. 



\section{Discussion}\label{sec:DISC}
In this paper we have proposed formula \ref{eq:Smatrix} to translate CFT correlators into flat space scattering amplitudes.  The proposal has been tested in simple $(2+1)$-dimensional examples. We have obtained low-point functions and blocks fixed by global BMS$_3$ invariance from low-point functions and global conformal blocks in CFT$_2$'s. We have also understood how non-trivial scattering processes can arise from more complicated conformal correlators. In particular, we have shown how the scattering of light particles against conical deficit geometries in flat space-times arises from the flat limit of conformal correlators in CFT$_2$ states different from the vacuum. 

There are some interesting lines of future research which we will describe briefly and speculatively here.

\noindent\textbf{Flat limit of Ward identities: }
In an upcoming paper \cite{AUXEHCR} we will explore how BMS ward identities arise from a flat limit of Ward identities in a conformal field theory. In dimensions higher than three ($D>3$), the Ward identities correspond to Weinberg soft theorems\cite{Weinberg:1965nx} involving soft radiative modes of the gravitational field \cite{He:2014laa,Strominger:2014pwa}. In three dimensions, we expect that the resulting flat limit results in the scattering of light particles over geometries containing boundary gravitons. 

\noindent\textbf{Non-trivial scattering in higher dimensions: } 
Formula \ref{eq:Smatrix} is in principle valid in higher dimensions, and can be adapted to asymptotically flat geometries containing black holes. It is tempting to state that the S-matrix amplitude of particles propagating in black hole geometries can be studied using conformal correlators through a flat limit. Studying such topics, and quantum corrections to formula \ref{eq:Smatrix}, could lead to new insights concerning information loss in Schwarzschild black holes. There are several complications to this logic. The main one is that asymptotically flat black holes can only arise as a flat limit of AdS geometries containing ``little" black holes, with a radius that would not scale with the AdS length scale. It is not clear (to us) how one can compute conformal correlators associated to such background. 

\noindent\textbf{BMS$_3$ blocks: }
In this work, we have compared results from the flat limit of global conformal correlators to kinematic invariants of a theory with BMS$_3$ symmetry. However, we have only required global BMS invariance, and so the results obtained in section \ref{sec:SMATRIX} are really fixed by Poincar\'e invariance. We believe it would be interesting to understand flat limits of Virasoro blocks, which should result in full BMS$_3$ blocks. It is not immediately obvious what such blocks correspond to in the language of asymptotically flat geometries, and so it is also interesting to understand the holographic picture associated to such objects. 

\noindent\textbf{Diagnostics of bulk locality in AdS/CFT: }
The formula we have studied in this paper can be understood as a way to carefully extract the contribution to a conformal correlator coming from a kinematic singularity, when the operators are located at configurations involving insertions \ref{eq:insertion} that conserve total energy-momentum.  The authors of \cite{Maldacena:2015iua} have explored such poles in the case of null momenta ($\omega_i=\mp k_i$). Some singularities can only be understood as Landau poles in the bulk, representing a scattering event between massless particles that conserve total momentum. The existence of such singularities is a clear diagnostic of bulk locality. We believe that the existence of singularities associated to scattering events involving external particles with time-like momenta is worth exploring. If such singularities happen when there is no clear singularity in the boundary, one would be tempted to think of this as a further diagnostic of bulk locality.


\acknowledgments
I am particularly grateful to Charles Rabideau for very useful discussions throughout the duration of the project. I also thank Tarek Anous, Felix Haehl, Per Kraus, Dominik Neuenfeld, Eric Perlmutter, Grant Salton and Allic  Sivaramakrishnan for their advice.

My work is supported in part by the Natural Sciences and Engineering Research Council of Canada  and the “It From Qubit” collaboration grant from the Simons Foundation. 


\appendix

\section{BMS$_3$ descendant operators}\label{app:desc}
Descendants of the state $\vert M,s\rangle$ defined in section \ref{sec:UniRep} are obtained by acting with generators of super-rotations $L_n$. In terms of descendant operators, these states correspond to  insertions  in points in super-momentum space with non-trivial values of $p^{\vert n \vert>1}$, which can be obtained by acting with finite generators of super-rotations on the operators in the rest frame
\be
\Phi(p) = U^{-1} \Phi(p_0) U\, , \quad \text{with} \quad U=\text{Exp}\left(   \sum_n \omega_n  L_n  \right)\, .
\ee
Here, the parameters $\omega_n$ parametrize the Fourier expansion of a super-rotation vector field on the circle $\omega(\phi)\partial_{\phi}$. Following a similar logic to the one used in section \ref{sec:StateOp}, one can derive the action of the generators of the symmetry algebra on descendant operators. As explained clearly in \cite{Campoleoni:2016vsh}, one can find
\be
[M_n,\Phi(p) ]= p^{(n)}  \Phi(p) \, .
\ee
The new super-momentum reads
\be
p(\phi)=\sum_n  p^{(n)} e^{i n \phi} -{{c_M}\over{24}}= {1\over{(f'(\phi))^2}} \left[  p_0+{{c_M}\over{12}} \{ f,\phi \}    \right]\, ,
\ee
where the function $f(\phi)$ parametrizes the super-rotation implemented by $U$, and it is related to the vector field $\omega(\phi)\partial_{\phi}$ as
\be
\int^{f(\phi_0)}_{\phi_0} {{d\phi}\over{\omega(\phi)}}=1\, .
\ee


\section{Amplitude with massless external particles}\label{app:massless}
In this appendix we implement the flat limit of a conformal correlator involving four operators with conformal dimensions $\Delta_i ={\cal O}(1)$. We start with formula \ref{eq:mellin}, which we repeat here
\be
\langle {\cal O}(x_1)...{\cal O}(x_n) \rangle = \int [d\gamma]\,  {\cal M}(\gamma_{ij}) \prod_{1\leq i\leq j\leq n}  \Gamma(\gamma_{ij}) \left(x_{ij}^2\right)^{-\gamma_{ij}}\, .
\ee
The operators appearing here have to be inserted at locations \ref{eq:insertion}, and then the correlator has to be smeared as in equation \ref{eq:Smatrix}. Replacing the Mellin variables by the scaling $\gamma_{ij}=l^2 \delta_{ij}$ yields the following equation for the kinematic factors appearing in the Mellin transform
\be
\begin{split}
\prod_{a<b}& \left( x_{ab}^2 \right)^{-\delta_{ab}}\Gamma(\gamma_{ab})
=
l^{-6} \left( {{2\pi}\over{\delta_{24}}} {{2\pi}\over{\delta_{34}}} {{-2\pi}\over{\delta_{24}+\delta_{34}}}   \right) e^{ F(t_i,\delta_{ij}) } \left( {{\delta_{34}^2}\over{(-\delta_{24}-\delta_{34})^2}} {{     \sin^2{{\phi_{23}}\over 2} \sin^2{{\phi_{14}}\over 2}          }\over{\sin^4{{{\phi_{12}}\over 2}}}}\right)^{\Delta}  e^{l^2 S(\delta_{ij})} 
\end{split}
\ee
with 
\be
S(\delta_{ij})=\delta_{24} \log\left( {{\delta_{24}^2}\over{(-\delta_{24}-\delta_{34})^2}} {{    \sin^2 {{\phi_{14}}\over 2}       \sin^2 {{\phi_{23}}\over 2}     }\over{     \sin^2 {{\phi_{13}}\over 2}       \sin^2 {{\phi_{24}}\over 2}        }} \right)
+\delta_{34} \log\left( {{\delta_{34}^2}\over{(-\delta_{24}-\delta_{34})^2}} {{    \sin^2 {{\phi_{23}}\over 2}       \sin^2 {{\phi_{14}}\over 2}     }\over{     \sin^2 {{\phi_{12}}\over 2}       \sin^2 {{\phi_{34}}\over 2}        }} \right)\, .
\ee
and
\be
\begin{split}
F(t_i,\delta_{ij}) &= {1\over 4} \delta_{24} \left(  {{t_{13}^2}\over{\sin^2{{\phi_{13}}\over 2}}}-{{t_{23}^2}\over{\sin^2{{\phi_{23}}\over 2}}} -{{t_{14}^2}\over{\sin^2{{\phi_{14}}\over 2}}} +{{t_{24}^2}\over{\sin^2{{\phi_{24}}\over 2}}}\right) \\
&+{1\over 4} \delta_{34}\left(  {{t_{12}^2}\over{\sin^2{{\phi_{12}}\over 2}}}  -{{t_{23}^2}\over{\sin^2{{\phi_{23}}\over 2}}}  -{{t_{14}^2}\over{\sin^2{{\phi_{14}}\over 2}}}+{{t_{34}^2}\over{\sin^2{{\phi_{34}}\over 2}}} \right)\, .
\end{split}
\ee
In the large $l$ limit, we can perform the integral over the Mellin variables using a stationary phase approximation. If we try to solve the condition $d S/d\delta_{24}=dS/d\delta_{34}=0$, we find a degenerate saddle at
\be
\delta_{24}=\delta_{24}^{*}=-\delta_{34} {{  \sin{{\phi_{13}}\over 2}  \sin{{\phi_{24}}\over 2}    }\over{  \sin{{\phi_{12}}\over 2}  \sin{{\phi_{34}}\over 2} }}\, .
\ee
We will thus consider a saddle point approximation for the integral over $\delta_{24}$, and leaving a reminding integral over $\delta_{34}$. The approximation reads
\be
\int d\gamma_{24} \, g(\delta_{24}) e^{l^2 S(\delta_{ij})}\sim l^2 g(\delta_{24}^*) \sqrt{{2\pi}\over{l^2 \partial_{\delta_{24}}^2S(\delta_{24}^*)}} e^{l^2 S(\delta^*_{ij})}
\ee
where $g$ contains the Mellin amplitude and the non-divergent part of the prefactors.  The on-shell action $S(\delta^*_{ij})$ vanishes. We thus have, in terms of the Mellin amplitude
\be
\begin{split}
{\cal S}&\sim l^3 \int d\delta_{34} \left[\prod_i \omega_i\int dt_i \, e^{i \omega_i t_i}\right] 
{\cal M}( \gamma_{24}=l^2 \delta_{24}^*, \gamma_{34}=l^2 \delta_{34} ) \\
&  l^{-6} \left( {{2\pi}\over{\delta^*_{24}}} {{2\pi}\over{\delta_{34}}} {{-2\pi}\over{\delta^*_{24}+\delta_{34}}}   \right)  \sqrt{{2\pi}\over{\partial_{\delta_{24}}^2S(\delta_{24}^*)}}     \left( {{\delta_{34}^2}\over{(-\delta_{24}^*-\delta_{34})^2}} {{     \sin^2{{\phi_{23}}\over 2} \sin^2{{\phi_{14}}\over 2}          }\over{\sin^4{{{\phi_{12}}\over 2}}}}\right)^{\Delta}  e^{ F(t_i,\delta_{34})} 
\end{split}
\ee
The integrals over the times $t_i$ are either Gaussian or Dirac deltas. They can be performed explicitly to obtain
\be
\begin{split}
{\cal S}\sim& \delta^{3}\left(\sum_i p_i \right) 
\int d\alpha \, e^{\alpha} \alpha \, {\cal M}\left( \gamma_{24}=l^2 \delta_{24}^*, \gamma_{34}=l^2 {{s_{34}}\over{4\alpha}}\right)\, .
\end{split}
\ee
where we have defined $s_{ij}=2p_i\cdot p_j$, and we have also changed variables for $\delta_{34}$ so the result looks exactly like the one derived in \cite{Fitzpatrick:2011hu}. This is the formula quoted in the main text. If one now uses the explicit expression for the Mellin amplitude of a conformal block, this formula reduces to the result \ref{eq:MasslessFinal}. We wont show the explicit computation here, as this was proven clearly in \cite{Fitzpatrick:2011hu}.


\section{Amplitude with massive external particles}\label{app:massive}

Unlike the massless case, here the Mellin constraints demand the scaling $\gamma_{ij}=l \delta_{ij}$. Otherwise the computation is very similar. Expanding the integrand at large $l$ we can approximate the result by a saddle point approximation. This yields two extremum equations that yield non-degenerate solutions for $\delta_{24}$ and $\delta_{34}$. The four integrals over $t_i$ result in four delta functions. Three of them will correspond to conservation of three-momenta, while the last one will fix the exchanged momentum in the block. The structure of the integral reads
\be
{\cal S}\sim l^{-4} \prod_i \int dt_i \, e^{i \omega_i t_i} \int [d\delta]\,  g(s_{ab},m_a,\delta_{ab}) e^{l S(\delta_{ab})} e^{i \sum_j t_j g_j(s_{ab},m_a,\delta_{ab})}\, .
\ee
Here, both $g(s_{ab},m_a,\delta_{ab})$ and  $S(\delta_{ab})$ contains dependence on the structure of the Mellin amplitude. The stationary phase approximation  requires us to evaluate the integrand at the saddle $\delta_{ab}^{*}$ obeying
\be\label{eq:Cond1M}
\partial_{\delta_{24}}S(\delta_{ab})\vert_{\delta_{ab}^{*}}=\partial_{\delta_{34}}S(\delta_{ab})\vert_{\delta_{ab}^{*}}=0\, .
\ee
The integrals over times $t_i$ are Dirac deltas of the form
\be\label{eq:Cond2M}
\prod_i \delta\left(  \omega_i +g_i(s_{ab},m_a,\delta_{ab})   \right)\, .
\ee
The conditions imposed by the distributions \ref{eq:Cond2M} can be used to simplify the saddle point equations \ref{eq:Cond1M}, which result in the following saddle
\be\label{eq:saddle}
\delta_{ab}^*={{m_a m_b+p_a\cdot p_b}\over{\sum_i m_i}}\, .
\ee
This is precisely the point in Mellin space relevant for the proposal in \cite{Paulos:2016fap}. We conclude that the formula \ref{eq:Smatrix} roughly reduces to the proposal in \cite{Paulos:2016fap} in the case of massive external particles.

Evaluating the Mellin integrals at the saddle \ref{eq:saddle}, reorganizing the arguments of the remaining Dirac delta functions, and cancelling the large $l$ divergences with the pre-factors in the proposed formula  for the S-matrix \ref{eq:Smatrix}, one obtains the expected result derived using  global BMS$_3$ symmetry in equation \ref{eq:block}
\be
{\cal S}\sim \delta^{(3)}\left(   \sum_{i} p_i   \right) \delta\left(  s+M^{\prime 2}    \right)\, ,
\ee
where the momenta appearing in this expression are time-like, obeying $p_i^2=-m_i^2$. 


\section{Scattering on a flat cone revisited}\label{app:Gcone}
In this appendix we solve the propagator $G(x',x)$ obeying the defining differential equation 
\be
\left(\nabla^2+m^2  \right) G(x,x')={1\over{\sqrt{-g}}} \delta^{(3)}(x-x')\, ,
\ee
with single-valuedness condition $G(\phi+2\pi)=G(\phi)$, on the cone geometry
\be
ds^2=- dt^2 +dr^2 +r^2\alpha^2 d\phi^2\, .
\ee
Some of the calculations shown here are in some ways similar to the ones in \cite{Moreira:1995he}. We first look for the eigenfunctions of the d'Alembertian
\be
\left( \nabla^2-m^2\right)\psi(x)=E\, \psi(x)\, .
\ee
The solutions that are well behaved at $r=0$ read
\be\label{eq:PsiCone}
\psi(x)= {1\over{\sqrt{2\pi}}}J_{{{\vert\nu\vert}\over{\alpha}}}(\kappa r) e^{i(-w  t + \nu \phi)}\, ,
\ee
with  eigenvalue $E=w^2-\kappa^2 -m^2$. Single valuedness of these solutions as $\phi\sim\phi+2\pi$ implies $\nu=n\in \mathbb{Z}$. We now use the following two equations (First one is an integral formula of the Bessel functions, and second one is Poisson's summation formula applied to the delta distribution)
\be
\begin{split}
\int_0^{\infty} d\kappa \, \kappa \, J_{\nu}(\kappa r)  J_{\nu}(\kappa r')     &=  {1\over r}\delta(r-r')\, ,\\
\sum_{n=-\infty}^{\infty}\delta(\theta+2\pi n)&={1\over {2\pi}}  \sum_{n=-\infty}^{\infty}e^{i n \theta}\, ,
\end{split}
\ee
to write an expression for the delta function as an integral over eigenfunctions of the d'alembertian. 
\be
{\cal I}=\sum_{n=-\infty}^{\infty} \int_{-\infty}^{\infty}dw\int_{0}^{\infty}d\kappa\, \kappa \, \psi_{n,\kappa,w}(x)\psi_{-n,\kappa,-w}(x')={1\over{\sqrt{g}}} \delta^{(3)}(x-x')\, .
\ee
We can now use the eigenvalue equation to note
\be
\begin{split}
\left( \nabla_x^2+m^2\right){\cal I}=\sum_{n=-\infty}^{\infty} \int_{-\infty}^{\infty}d\omega \int_{0}^{\infty}d\kappa\, \kappa \, E_{\kappa,w} \psi_{n,\kappa,w}(x)\psi_{-n,\kappa,-w}(x')\, .
\end{split}
\ee
We can thus define the propagator by dividing the integrand over the eigenvalue $E_{\kappa,\omega}$
\be
G(x,x')= \sum_{n=-\infty}^{\infty} \int_{-\infty}^{\infty}d\omega \int_{0}^{\infty}d\kappa\, \kappa \,{1\over{E_{\kappa,w}+i\epsilon}} \psi_{n,\kappa,w}(x)\psi_{-n,\kappa,-w}(x')\, ,
\ee
Where the $\epsilon$ prescription has been chosen so that the propagator is the Feynman one. Explicitly, this formula reads
\be
\begin{split}
G(x,x')=& \sum_{n=-\infty}^{\infty}  e^{i n (  \phi-  \phi')}  \int_{0}^{\infty}d\kappa\, \kappa \,  J_{\left\vert{n\over{\alpha}}\right\vert }(\kappa r) J_{\left\vert{n\over{\alpha}}\right\vert }(\kappa r') \int_{-\infty}^{\infty}dw \, {{ e^{-iw (t - t' )}}\over{w^2-\kappa^2-m^2+i\epsilon}}    \, .
\end{split}
\ee
The integral over $\omega$ can be performed by closing the contour along large imaginary values of $\omega$ if $t<t'$ or negatively large values if $t>t'$. Here, we will not perform the integral and continue with the Fourier transform suggested in the main text. 
\be
G(p,p')=\int d^3 x \, e^{i \left(  -\omega t + k r \cos(\phi-\chi)  \right)}  \int d^3 x' \, e^{i \left(  -\omega' t' + k' r' \cos(\phi'-\chi')  \right)} \, G(x,x')\, .
\ee
In order to perform the integrals over the space-time points, we will make use of the following representation of the plane-waves 
\be
e^{i \left(  -\omega t + k r \cos(\phi-\chi)  \right)} =e^{-i \omega t} \sum_n e^{i {{\pi}\over 2} \vert n \vert} e^{i n (\phi-\chi)} J_{\vert n \vert} ( k  r )\, .
\ee
The resulting computation reads
\be
\begin{split}
G(p,p')&=\int d^3x  \int d^3x' \sum_{n,s,s'} i^{\vert s \vert} i^{\vert s' \vert} e^{i \phi (s+n)}e^{i \phi' (s'-n)}  e^{-i \omega t-i\omega' t'}  J_{\vert s \vert} ( k  r ) J_{\vert s' \vert} ( k'  r' )\\
&\times  e^{-i s \chi}e^{-i s' \chi' } \int_{0}^{\infty}d\kappa\, \kappa \,  J_{\left\vert{n\over{\alpha}}\right\vert }(\kappa r) J_{\left\vert{n\over{\alpha}}\right\vert }(\kappa r')  \int_{-\infty}^{\infty}dw\, {{ e^{-i w (t - t' )}}\over{w^2-\kappa^2-m^2+i\epsilon}} 
\end{split}
\ee
The integrals over times and angles are simple delta functions which simplify the computation
\be
\begin{split}
G(p,p')&=\delta(\omega+\omega')\int rdr  \int r'dr' \sum_{n}e^{i n (\chi-\chi')} e^{i \pi \vert n \vert}     J_{\vert n \vert} ( k  r ) J_{\vert n \vert} ( k'  r' )\\
&\times   \int_{0}^{\infty}d\kappa\, \kappa \,  J_{\left\vert{n\over{\alpha}}\right\vert }(\kappa r) J_{\left\vert{n\over{\alpha}}\right\vert }(\kappa r') \, {{ 1 }\over{\omega^2-\kappa^2-m^2+i\epsilon}} 
\end{split}
\ee
The integral over $\kappa$ can be performed by using formula 1 in section 6.541 of \cite{ListIntegrals}. 
\be
 \int_{0}^{\infty}d\kappa\, \, {{  \kappa \,  J_{\left\vert{n\over{\alpha}}\right\vert }(\kappa r) J_{\left\vert{n\over{\alpha}}\right\vert }(\kappa r') }\over{\omega^2-\kappa^2-m^2+i\epsilon}} =  i {{\pi}\over 2} \begin{cases}
       J_{\left\vert{n\over{\alpha}}\right\vert }\left( \kappa_* r \right) H^{(2)}_{\left\vert{n\over{\alpha}}\right\vert } \left(\kappa_* r' \right)& \text{if: } r<r'\\
       J_{\left\vert{n\over{\alpha}}\right\vert }\left(\kappa_*  r'  \right) H^{(2)}_{\left\vert{n\over{\alpha}}\right\vert } \left(\kappa_*  r \right)& \text{if: } r>r'
    \end{cases}   \, ,
\ee
where we have defined $\kappa_*=\sqrt{\omega^2-m^2}$  and taken the $\epsilon\rightarrow 0 $ limit after performing the integral. The remaining radial integrals are
\be
\begin{split}
G(p,p')&= i {{\pi}\over 2} \delta(\omega+\omega')\sum_{n} e^{i n (\chi-\chi')} e^{i \pi \vert n \vert}   \\
&\times \left( \int_{r'>r} dr dr'  \, r  J_{\vert n \vert} ( k  r )  J_{\left\vert{n\over{\alpha}}\right\vert }\left( \kappa_* r \right)  \,   r'   J_{\vert n \vert} ( k'  r' ) H^{(2)}_{\left\vert{n\over{\alpha}}\right\vert } \left(\kappa_* r' \right)     \right.   \\
& \left.  \quad  +    \int_{r>r'} dr dr' \, r  J_{\vert n \vert} ( k  r )  H^{(2)}_{\left\vert{n\over{\alpha}}\right\vert }\left( \kappa_* r \right)    \, r'    J_{\vert n \vert} ( k'  r' ) J_{\left\vert{n\over{\alpha}}\right\vert } \left(\kappa_* r' \right)      \right)
\end{split}
\ee
relabeling the radial integrals in the second line yields
\be\label{eq:auxapp1}
\begin{split}
G(p,p')&= i {{\pi}\over 2} \delta(\omega+\omega')\sum_{n} e^{i n (\chi-\chi')} e^{i \pi \vert n \vert}   \int_{r'>r} dr dr'   \\
&\times   r r'\,   \left[  J_{\vert n \vert} ( k  r )  J_{\vert n \vert} ( k'  r' ) + J_{\vert n \vert} ( k'  r )  J_{\vert n \vert} ( k  r' )      \right] J_{\left\vert{n\over{\alpha}}\right\vert }\left( \kappa_* r \right)    H^{(2)}_{\left\vert{n\over{\alpha}}\right\vert } \left(\kappa_* r' \right)   
\end{split}
\ee
Before continuing, we will add and substract the propagator for $\alpha=1$. The addition will result on the standard result for scattering of particles in Minkowski space, resulting in a delta function conserving momentum. The substraction will be mixed with the terms in \ref{eq:auxapp1}. This yields
\be
\begin{split}
G(p,p')-\delta^{(3)}(p+p')&=\delta(\omega+\omega')\sum_{n} e^{i n (\chi-\chi')} e^{i \pi \vert n \vert}   \int_{r'>r} dr dr'  \, r r' \\
&\times   \left[  J_{\vert n \vert} ( k  r )  J_{\vert n \vert} ( k'  r' ) + J_{\vert n \vert} ( k'  r )  J_{\vert n \vert} ( k  r' )      \right]   \\
&\times \left[   J_{\left\vert{n\over{\alpha}}\right\vert }\left( \kappa_* r \right)    H^{(2)}_{\left\vert{n\over{\alpha}}\right\vert } \left(\kappa_* r' \right)    -J_{\left\vert{n}\right\vert }\left( \kappa_* r \right)    H^{(2)}_{\left\vert{n}\right\vert } \left(\kappa_* r' \right)            \right]
\end{split}
\ee
We can now perform the radial integrals as follows. We first re-write the integration limits as
\be
  \int_{r'>r} dr dr'  = \lim_{\Lambda\rightarrow \infty}\int_0^{\Lambda} \int_{r'=r}^{\Lambda} dr dr'   
\ee
we now replace $r=\Lambda \tilde{r}$ and $r'=\Lambda \tilde{r}'$ so we have
\be
\begin{split}
\lim_{\Lambda\rightarrow \infty} &\Lambda^4 \int_0^{1}d\tilde{r} \int_{\tilde{r}}^1 d\tilde{r}'     \tilde{r} \tilde{r}'\,   \left[  J_{\vert n \vert} ( \Lambda k  \tilde{r} )  J_{\vert n \vert} ( \Lambda k'  \tilde{r}' ) + J_{\vert n \vert} ( \Lambda k' \tilde{r} ) J_{\vert n \vert} ( \Lambda k  \tilde{r}' )      \right] \\
&\times \left[  J_{\left\vert{n\over{\alpha}}\right\vert }\left( \Lambda \kappa_* \tilde{r} \right)    H^{(2)}_{\left\vert{n\over{\alpha}}\right\vert } \left(\Lambda \kappa_* \tilde{r}' \right) -   J_{\left\vert{n}\right\vert }\left( \Lambda \kappa_* \tilde{r} \right)    H^{(2)}_{\left\vert{n}\right\vert } \left(\Lambda \kappa_* \tilde{r}' \right)      \right]\, .
\end{split}
\ee
The integral can be performed easily by making use of the asymptotic form of the Bessel functions. Note that this is allowed, as the functions themselves do not contribute when their arguments vanish. This can also be said about the Bessel $H^{(2)}$, as the regime of integration avoids its divergence at vanishing argument. The formulas we use are
\be
\begin{split}
J_{ \nu  }( \Lambda x) &\approx \sqrt{{2\over{\pi \Lambda x}}}\,  \cos\left(   \Lambda x - \nu {{\pi}\over 2} - {{\pi}\over 4}  \right)\, , \\
H^{(2)}_{ \nu  }( \Lambda x) &\approx \sqrt{{2\over{\pi \Lambda x}}}\,  e^{i \left(  - \Lambda x + \nu {{\pi}\over 2} + {{\pi}\over 4}  \right)  }\, .
\end{split}
\ee
The resulting radial integrals turn into simple delta functions. In the end, we can write
\be
\begin{split}
G(p,p')&=\delta^{(3)}(p+p')-   \delta(\omega+\omega'){1\over{k}}\delta(k-k')\sum_{n} e^{i n (\chi-\chi'+\pi)}  \left( e^{i \pi \vert n\vert    \left(   {1\over {\alpha}}-1     \right)  } -1\right)  \\
&=\delta^{(3)}(p+p')+\delta(\omega+\omega'){1\over{k}}\delta(k-k')  f(\chi,\chi')
\end{split}
\ee
where we have defined
\be
f(\chi,\chi')= i {{\sin{{\pi}\over {\alpha}}}\over{  \cos{{\pi}\over{\alpha}}    - \cos(\chi-\chi')     }}\, ,
\ee
and the momentum vectors are on-shell. This matches the non-relativistic result found in  \cite{tHooft:1988qqn,Deser:1988qn}, and it is an altenate derivation to the relativistic calculation presented in \cite{Moreira:1995he}. For a pictorial interpretation of this result, see figure \ref{fig:conepic}.


\section{Propagator in an AdS$_3$ conical deficit}\label{app:AdScone}
Here we derive formula   \ref{eq:2ptConeCFT}       for the holographic two-point function in a conical deficit AdS$_3$ geometry with line element 
\be\label{eq:AdSConeGeom}
ds^2={{\alpha^2 l^2}\over{\cos^2\rho}} \left(  {{1}\over{\alpha^2}}d\rho^2 -d\tau^2 +\sin^2\rho d\phi^2      \right)\, .
\ee
We start as we did in the flat case, by studying the propagator in a defect geometry. It will be useful to consider the following eigenvectors of the d'alembertian
\be
\left( {1\over{\alpha^2}}\cot^2\rho \partial^2_{\phi}+\cot\rho \partial_{\rho} +\cos^2\rho \partial^2_{\rho} -{1\over{\alpha^2}}\cos^2\rho \partial^2_{\tau} - \Delta(\Delta-2)    \right) \psi =E \cos^2\rho  \, \psi \, .
\ee
The solutions that do not diverge at the boundary  read
\be
\begin{split}
\psi_{E,\omega,\nu} &= \left(   \sin\rho \right)^{{{\vert \nu \vert}\over{\alpha}}} \left(  \cos\rho  \right)^{1+\sqrt{1+m^2}}    e^{-i\alpha\omega \tau} e^{i \nu \phi} \\
\times {}_2 F_1 &\left(  {{1+\sqrt{1+m^2}-\sqrt{\omega^2-E}}\over 2} + {{\vert \nu\vert }\over{2\alpha}}  , {{1+\sqrt{1+m^2}+\sqrt{\omega^2-E}}\over 2} + {{\vert \nu\vert }\over{2\alpha}} ; 1+\sqrt{1+m^2} \vert \cos^2\rho \right)
\end{split}
\ee
In order to remove the branch cuts of the hypergeometric function at the center of global AdS, we require
\be
{{1+\sqrt{1+m^2}-\sqrt{\omega^2-E}}\over 2} + {{\vert \nu\vert }\over{2\alpha}}=-\kappa\in\mathbb{Z}
\ee
such that
\be
\begin{split}
\psi_{E,\omega,\nu} =& \left(   \sin\rho \right)^{{{\vert \nu \vert}\over{\alpha}}} \left(  \cos\rho  \right)^{1+\sqrt{1+m^2}}   e^{-i\alpha\omega \tau} e^{i \nu \phi}\\
&\times {}_2 F_1\left(  -\kappa  ,\kappa+1+\sqrt{1+m^2} + {{\vert \nu\vert }\over{\alpha}} ; 1+\sqrt{1+m^2} \vert \cos^2\rho \right)
\end{split}
\ee
This can be written as a Jacobi polynomial (ignoring overall constants)
\be
\psi_{E,\omega,\nu} = \left(   \sin\rho \right)^{{{\vert \nu \vert}\over{\alpha}}} \left(  \cos\rho  \right)^{1+\sqrt{1+m^2}}   e^{-i\alpha\omega \tau} e^{i \nu \phi} P_{\kappa}^{\left(  {{\vert \nu \vert}\over{\alpha}}, \sqrt{1+m^2}  \right)}\left(  \cos2\rho  \right)
\ee
Single-valuedness in the angle coordinate implies $\nu=n\in\mathbb{Z}$. The Jacobi polynomials obey the following identity
\be
\sum_{p=0}^{\infty} {{p! (a+b+2p+1)\Gamma(a+b+p+1)}\over{\Gamma(a+p+1)\Gamma(b+p+1)}}  P_p^{(a,b)}(x)  P_p^{(a,b)}(y) =  {{2^{a+b+1} \delta(x-y)}\over{ ((1-x)(1-y))^{{a\over 2}}   ((1+x)(1+y))^{{b\over 2}}}}  
\ee
So if we define  $\Phi_{\kappa, \omega,n}=C_{\kappa,n}\psi_{\kappa,\omega,n}$ with
\be
C_{\kappa,n}=\sqrt{{{2 \kappa ! \left({{|n|}\over{\alpha}}+\sqrt{1+m^2}+2\kappa+1\right)\Gamma\left({{|n|}\over{\alpha}}+\sqrt{1+m^2}+\kappa+1\right)}\over{\alpha \Gamma\left(\sqrt{1+m^2}+\kappa+1\right)\Gamma\left({{|n|}\over{\alpha}}+\kappa+1\right)}}}
\ee
and also
\be
E=\omega^2-\omega_{\kappa,n}^2\, , \quad \text{with}\quad \omega_{\kappa,n}={1+\sqrt{1+m^2}}+{{\vert n \vert}\over{\alpha}}+ 2\kappa
\ee
we have
\be\begin{split}
\left( \nabla^2-m^2\right)\sum_{p=0}^{\infty}{1\over{\omega^2-\omega_{\kappa,n}^2}} \Phi_{\kappa, \omega, n}^{*}(x')\Phi_{\kappa, \omega, n}(x) &=\cos^2 \rho  \sum_{\kappa =0}^{\infty}{E\over{\omega^2-\omega_{\kappa,n}^2}}  \, \Phi_{\kappa, \omega, n}^{*}(x')\Phi_{\kappa, \omega, n}(x) \\
&= {1\over {\alpha}}{{\cos^3\rho}\over{\sin\rho}} \delta(\rho-\rho') e^{i n (\phi-\phi')} e^{-i \alpha \omega(t-t')}
\end{split}
\ee
By further summing over $n$ and integrating over $\omega$ we have
\be
\left( \nabla^2-m^2\right)\int {{d\omega}\over{2\pi}}\sum_n\sum_{p=0}^{\infty}{1\over{\omega^2-\omega_{\kappa,n}^2}} \Phi_{\kappa, \omega, n}^{*}(x')\Phi_{\kappa, \omega, n}(x) ={1\over {\alpha^2}}{{\cos^3\rho}\over{\sin\rho}} \delta(\rho-\rho') \delta(t-t')\delta(\phi-\phi')
\ee
so we conclude that the propagator reads
\be
G(X,X')=\int {{d\omega}\over{2\pi}}\sum_n\sum_{\kappa =0}^{\infty}{1\over{\omega^2-\omega_{\kappa,n}^2}} \Phi_{\kappa, \omega, n}^{*}(t',\phi',\rho')\Phi_{\kappa, \omega, n}(t,\phi,\rho) \, .
\ee
We are now interested in the boundary two point function, so we look at the propagator close to $\rho,\rho'\sim\pi/2$. The eigenfunctions read
\be
\Phi_{\kappa, \omega,n}\rightarrow C_{\kappa,n}   \left(  \cos\rho  \right)^{1+\sqrt{1+m^2}}   e^{-i\alpha\omega \tau} e^{i n \phi}  (-1)^{\kappa}  {{(1+\sqrt{1+m^2})_{\kappa}}\over{\kappa !}}
\ee
and so
\be
\begin{split}
K(X,X')&=\lim_{\rho,\rho'\rightarrow {{\pi}\over 2}}{1\over{(\cos\rho)^{1+\sqrt{1+m^2}}}}{1\over{(\cos\rho')^{1+\sqrt{1+m^2}}}} G(X,X')\\
&=\int {{d\omega}\over{2\pi}}\sum_n\sum_{\kappa=0}^{\infty}{1\over{\omega^2-\omega_{\kappa,n}^2}}    e^{i n (\phi-\phi')}  e^{-i\alpha\omega (\tau-\tau')}  \left(      {{(1+\sqrt{1+m^2})_{\kappa}}\over{\kappa !}}   \right)^2\\
&\times {{2 \kappa ! \left({{|n|}\over{\alpha}}+\sqrt{1+m^2}+2\kappa+1\right)\Gamma\left({{|n|}\over{\alpha}}+\sqrt{1+m^2}+\kappa+1\right)}\over{\alpha \Gamma\left(\sqrt{1+m^2}+\kappa+1\right)\Gamma\left({{|n|}\over{\alpha}}+\kappa+1\right)}}  \\
&\sim \sum_n\sum_{\kappa=0}^{\infty}    e^{i n (\phi-\phi')}  e^{-i\alpha\left( \Delta+{{\vert n \vert}\over{\alpha}}+ 2\kappa  \right) (\tau-\tau')}  
 {{\Gamma\left({{|n|}\over{\alpha}}+\Delta+\kappa\right)\Gamma\left(\Delta+\kappa\right)}\over{\Gamma\left({{|n|}\over{\alpha}}+\kappa+1\right)  \kappa ! }} \, .
\end{split}
\ee
This is the holographic derivation of the spectral decomposition formula \ref{eq:2ptConeCFT}. 


\section{Flat limit of CFT$_2$ defect two-point function}\label{app:flatlimitcone}
In this appendix we evaluate the flat limit of a two-point function in a CFT$_2$ state dual to a conical deficit in AdS$_3$. The formula we start with is
\be
{\cal S}\{ p_1,p_2 \} =\lim_{l\rightarrow\infty}\left[ \prod_{i=1}^2 \left(   {{m_i}\over {l k_i}}   \right)^{\Delta_i} \int dt_i \, e^{ i \omega_i t_i}\right] 
\sum_n\sum_{\kappa=0}^{\infty}    e^{i n (\chi_1-\chi_2)}  e^{- i\left(  \Delta+{{\vert n \vert}\over{\alpha}}+ 2\kappa \right) \tau_{12}} 
C_{\kappa,n}
\, ,
\ee
where we have defined
\be
\begin{split}
C_{n,\kappa}=  
 {{\Gamma\left({{|n|}\over{\alpha}}+\Delta+\kappa\right)\Gamma\left(\Delta+\kappa\right)}\over{\Gamma\left({{|n|}\over{\alpha}}+\kappa+1\right)  \kappa ! }}    \quad \text{and}\quad 
\tau_{12}=i\left( \cosh^{-1}{{\omega_1}\over{k_1}}-\cosh^{-1}{{\omega_2}\over{k_2}} \right)+{{t_{1}-t_2}\over l}  \, .
\end{split}
\ee
The integrals with respect to time can be reorganized in terms of the variables $t_{\pm}=t_1\pm t_2$,  such that
\be
 \prod_{i=1}^2  \int dt_1\, \int dt_2 \, e^{i \omega_1 t_1}e^{i \omega_1 t_2} ={1\over 2}\int dt_+ \, e^{i t_+{{\omega_1+\omega_2}\over 2}}\int dt_-  \, e^{i t_-{{\omega_1-\omega_2}\over 2}}
\, .
\ee
The integral over $t_+$ can be performed inmediately as the rest of the integrand does not depend on $t_+$. The result is a dirac delta function imposing conservation of the energy. The integral over $t_-$ seems to imply that the sum over $\kappa$ is dominated by large integers. To evaluate the sum explicitly, we re-write it as a hypergeometric function. The strategy is then to use an integral representation which allows for a saddle point approximation in the large $l$ limit. The sum reads
\be
\begin{split}
K&\equiv \sum_{\kappa=0}^{\infty}    e^{- i\left(  \Delta+{{\vert n \vert}\over{\alpha}}+ 2\kappa \right) \tau_{12}} 
C_{\kappa,n}  \\
&=
  e^{ -i \left(  \Delta+{{\vert n \vert}\over{\alpha}}   \right) \tau_{12}}    {{\Gamma\left(  \Delta  \right) \Gamma\left(  \Delta+ {{\vert n \vert}\over{\alpha}}  \right)    }\over{ \Gamma\left(  1+ {{\vert n \vert}\over{\alpha}}  \right) }} {}_2 F_1\left(   \Delta\, ,\, {{\vert n \vert}\over{\alpha}}+\Delta\, ; \, {{\vert n \vert}\over{\alpha}}+1\, \vert \,    e^{-2 i\tau_{12}}       \right)\, .
\end{split}
\ee
We now represent the hypergeometric function as a contour integral in the complex plane. We have
\be
\begin{split}
K=
  e^{ -i \left(  \Delta+{{\vert n \vert}\over{\alpha}}   \right) \tau_{12}}     \int_{-i\infty}^{i \infty} {{ds}\over{2\pi i}}     {{\Gamma\left(   s  \right)   \Gamma\left(  \Delta-  s  \right)    \Gamma\left(  {{\vert n \vert}\over{\alpha}}+\Delta- s  \right)  }\over{  \Gamma\left(  {{\vert n \vert}\over{\alpha}}+1- s  \right)    }}    \left(  -e^{-2 i\tau_{12}}       \right)^{-s}\, .
\end{split}
\ee
This integral representation is only valid if $\vert\text{arg}\left(   -e^{-2 i\tau_{12}}  \right) \vert<\pi$, which in our case is valid if we take the large $l$ limit after performing the integral over $s$. Note that this representation is very similar to the Melin representation of CFT correlators, and at the technical level, this calculation is very similar to the one found in appendices \ref{app:massive} and \ref{app:massless}. The integral in the large $l$ limit is dominated by a saddle point where $s$ scales as $s=l s'$. Replacing also $\Delta= m l$, we can write
\be
K={1\over l} e^{-2 m l}\int ds' e^{i t_-  \left(  2s'-m  \right)} g(s')    \,   e^{l f(s')}\, ,
\ee
where we have defined 
\be
\begin{split}
f(s')&=2 s' \log s' +2(m-s')\log(m-s') +(m-2s') \log {{k_1    \left(  \omega_2+\sqrt{\omega_2^2-k_2^2}    \right) }\over{k_2 \left(  \omega_1+\sqrt{\omega_1^2-k_1^2}    \right) }}\, , \\
g(s')&= (-1)^{  -  {{\vert n \vert}\over{\alpha}}   }    (m-s')^{  {{\vert n \vert}\over{\alpha}}  -1 } (s')^{-{{\vert n \vert}\over{\alpha}}-1}   \left(   {{k_1    \left(  \omega_2+\sqrt{\omega_2^2-k_2^2}    \right) }\over{k_2 \left(  \omega_1+\sqrt{\omega_1^2-k_1^2}    \right) }}    \right)^{  {{\vert n \vert}\over{\alpha}}  }  \, .
\end{split}
\ee
The integral can then be evaluated using a saddle point approximation. We look for saddle points obeying $\partial_{s'}f(s')=0$.  The saddle is located at
\be
s'_*=   {{m }\over{2(k_2\omega_1+k_1\omega_2)}}   \left[   k_2\left(  \omega_1-\sqrt{\omega_1^2-k_1^2}  \right)        + k_1\left(  \omega_2+\sqrt{\omega_2^2-k_2^2}  \right)         \right]\, .
\ee
this yields
\be
K={1\over l} e^{-2 m l} l^{2m l} e^{i t_-  \left(  2s'_* -m  \right)} g(s'_*)   \sqrt{ {2\pi}\over{l f''(s'_*)}    } \,   e^{l f(s'_*)}\, .
\ee
Before writting this expression explicitly, we now perform the integrals over the time parameters $t_{\pm}$. These now result in dimple dirac deltas that then will allow for a great simplification of the final result. The integral over $t_+$ yields a delta function for conservation of the energy. The integral over $t_-$ is more complicated, but it constraints $k_2=k_1$. Using delta function identities, the result of these integrals simplifies to
\be
{1\over 2}\int dt_+ \, e^{i t_+{{\omega_1+\omega_2}\over 2}}\int dt_-  \, e^{i t_-    \left(   {{\omega_1-\omega_2}\over 2} + 2s'_* -m      \right)}   =  {{2\left(  \omega_1^2-k_1^2  \right)^{3\over 2}}\over{m \omega_1}}   \delta(\omega_1+\omega_2){1\over k_1}\delta(k_1-k_2)\, .
\ee
Using now $\omega_2=-\omega_1$ and $k_2=k_1$ in our saddle point approximation, an further using the on-shell condition for the momenta results in
\be
{\cal S}\{ p_1,p_2 \} \sim \delta(\omega_1+\omega_2){1\over k_1}\delta(k_1-k_2)  
\sum_n  e^{i n (\chi_1-\chi_2)}  (-1)^{  - {{\vert n \vert}\over{\alpha}}      }
\ee
Adding and substracting this result with $\alpha=1$ yields
\be
{\cal S}\{ p_1,p_2 \} \sim \delta^{(3)}(p_1+p_2)+\delta(\omega_1+\omega_2){1\over k_1}\delta(k_1-k_2)  
\sum_n  e^{i n (\chi_1-\chi_2)} \left[ (-1)^{  - {{\vert n \vert}\over{\alpha}}      }-(-1)^{  - {\vert n \vert}     }   \right]
\ee
And finally performing the sum over $n$ we obtain
\be
{\cal S}\{ p_1,p_2 \} \sim \delta^{(3)}(p_1+p_2)+\delta(\omega_1+\omega_2){1\over k_1}\delta(k_1-k_2)  
{{i\sin{{\pi}\over{\alpha}}}\over{\cos{{\pi}\over{\alpha}}-\cos\chi_{12}}}\, .
\ee
This is the result quoted in the main text, which matches the result concerning scattering of particles in asymptotically flat conical deficit geometries.


\bibliographystyle{JHEP}
\bibliography{refs}

\providecommand{\href}[2]{#2}\begingroup\raggedright\begin{thebibliography}{10}

\bibitem{tHooft:1999rgb}
G.~'t~Hooft, \emph{{The Holographic principle: Opening lecture}},
  \href{http://dx.doi.org/10.1142/9789812811585_0005}{\emph{Subnucl. Ser.} {\bf
  37} (2001) 72--100}, [\href{https://arxiv.org/abs/hep-th/0003004}{{\tt
  hep-th/0003004}}].

\bibitem{Susskind:1994vu}
L.~Susskind, \emph{{The World as a hologram}},
  \href{http://dx.doi.org/10.1063/1.531249}{\emph{J. Math. Phys.} {\bf 36}
  (1995) 6377--6396}, [\href{https://arxiv.org/abs/hep-th/9409089}{{\tt
  hep-th/9409089}}].

\bibitem{Maldacena:1997re}
J.~M. Maldacena, \emph{{The Large N limit of superconformal field theories and
  supergravity}}, \href{http://dx.doi.org/10.1023/A:1026654312961}{\emph{Int.
  J. Theor. Phys.} {\bf 38} (1999) 1113--1133},
  [\href{https://arxiv.org/abs/hep-th/9711200}{{\tt hep-th/9711200}}].

\bibitem{Giddings:1999jq}
S.~B. Giddings, \emph{{Flat space scattering and bulk locality in the AdS / CFT
  correspondence}},
  \href{http://dx.doi.org/10.1103/PhysRevD.61.106008}{\emph{Phys. Rev.} {\bf
  D61} (2000) 106008}, [\href{https://arxiv.org/abs/hep-th/9907129}{{\tt
  hep-th/9907129}}].

\bibitem{Gary:2009mi}
M.~Gary and S.~B. Giddings, \emph{{The Flat space S-matrix from the AdS/CFT
  correspondence?}},
  \href{http://dx.doi.org/10.1103/PhysRevD.80.046008}{\emph{Phys. Rev.} {\bf
  D80} (2009) 046008}, [\href{https://arxiv.org/abs/0904.3544}{{\tt
  0904.3544}}].

\bibitem{Giddings:1999qu}
S.~B. Giddings, \emph{{The Boundary S matrix and the AdS to CFT dictionary}},
  \href{http://dx.doi.org/10.1103/PhysRevLett.83.2707}{\emph{Phys. Rev. Lett.}
  {\bf 83} (1999) 2707--2710},
  [\href{https://arxiv.org/abs/hep-th/9903048}{{\tt hep-th/9903048}}].

\bibitem{Balasubramanian:1999ri}
V.~Balasubramanian, S.~B. Giddings and A.~E. Lawrence, \emph{{What do CFTs tell
  us about Anti-de Sitter space-times?}},
  \href{http://dx.doi.org/10.1088/1126-6708/1999/03/001}{\emph{JHEP} {\bf 03}
  (1999) 001}, [\href{https://arxiv.org/abs/hep-th/9902052}{{\tt
  hep-th/9902052}}].

\bibitem{Penedones:2010ue}
J.~Penedones, \emph{{Writing CFT correlation functions as AdS scattering
  amplitudes}}, \href{http://dx.doi.org/10.1007/JHEP03(2011)025}{\emph{JHEP}
  {\bf 03} (2011) 025}, [\href{https://arxiv.org/abs/1011.1485}{{\tt
  1011.1485}}].

\bibitem{Fitzpatrick:2011jn}
A.~L. Fitzpatrick and J.~Kaplan, \emph{{Scattering States in AdS/CFT}},
  \href{https://arxiv.org/abs/1104.2597}{{\tt 1104.2597}}.

\bibitem{Maldacena:2011nz}
J.~M. Maldacena and G.~L. Pimentel, \emph{{On graviton non-Gaussianities during
  inflation}}, \href{http://dx.doi.org/10.1007/JHEP09(2011)045}{\emph{JHEP}
  {\bf 09} (2011) 045}, [\href{https://arxiv.org/abs/1104.2846}{{\tt
  1104.2846}}].

\bibitem{Raju:2012zr}
S.~Raju, \emph{{New Recursion Relations and a Flat Space Limit for AdS/CFT
  Correlators}},
  \href{http://dx.doi.org/10.1103/PhysRevD.85.126009}{\emph{Phys. Rev.} {\bf
  D85} (2012) 126009}, [\href{https://arxiv.org/abs/1201.6449}{{\tt
  1201.6449}}].

\bibitem{Fitzpatrick:2011hu}
A.~L. Fitzpatrick and J.~Kaplan, \emph{{Analyticity and the Holographic
  S-Matrix}}, \href{http://dx.doi.org/10.1007/JHEP10(2012)127}{\emph{JHEP} {\bf
  10} (2012) 127}, [\href{https://arxiv.org/abs/1111.6972}{{\tt 1111.6972}}].

\bibitem{Fitzpatrick:2011dm}
A.~L. Fitzpatrick and J.~Kaplan, \emph{{Unitarity and the Holographic
  S-Matrix}}, \href{http://dx.doi.org/10.1007/JHEP10(2012)032}{\emph{JHEP} {\bf
  10} (2012) 032}, [\href{https://arxiv.org/abs/1112.4845}{{\tt 1112.4845}}].

\bibitem{Paulos:2016fap}
M.~F. Paulos, J.~Penedones, J.~Toledo, B.~C. van Rees and P.~Vieira, \emph{{The
  S-matrix bootstrap. Part I: QFT in AdS}},
  \href{http://dx.doi.org/10.1007/JHEP11(2017)133}{\emph{JHEP} {\bf 11} (2017)
  133}, [\href{https://arxiv.org/abs/1607.06109}{{\tt 1607.06109}}].

\bibitem{Gary:2009ae}
M.~Gary, S.~B. Giddings and J.~Penedones, \emph{{Local bulk S-matrix elements
  and CFT singularities}},
  \href{http://dx.doi.org/10.1103/PhysRevD.80.085005}{\emph{Phys. Rev.} {\bf
  D80} (2009) 085005}, [\href{https://arxiv.org/abs/0903.4437}{{\tt
  0903.4437}}].

\bibitem{Hamilton:2006az}
A.~Hamilton, D.~N. Kabat, G.~Lifschytz and D.~A. Lowe, \emph{{Holographic
  representation of local bulk operators}},
  \href{http://dx.doi.org/10.1103/PhysRevD.74.066009}{\emph{Phys. Rev.} {\bf
  D74} (2006) 066009}, [\href{https://arxiv.org/abs/hep-th/0606141}{{\tt
  hep-th/0606141}}].

\bibitem{Deser:1988qn}
S.~Deser and R.~Jackiw, \emph{{Classical and Quantum Scattering on a Cone}},
  \href{http://dx.doi.org/10.1007/BF01466729}{\emph{Commun. Math. Phys.} {\bf
  118} (1988) 495}.

\bibitem{tHooft:1988qqn}
G.~'t~Hooft, \emph{{Nonperturbative Two Particle Scattering Amplitudes in
  (2+1)-Dimensional Quantum Gravity}},
  \href{http://dx.doi.org/10.1007/BF01218392}{\emph{Commun. Math. Phys.} {\bf
  117} (1988) 685}.

\bibitem{Spinally:2000ii}
J.~Spinally, E.~R. Bezerra~de Mello and V.~B. Bezerra, \emph{{Relativistic
  quantum scattering on a cone}},
  \href{http://dx.doi.org/10.1088/0264-9381/18/8/311}{\emph{Class. Quant.
  Grav.} {\bf 18} (2001) 1555--1566},
  [\href{https://arxiv.org/abs/gr-qc/0012103}{{\tt gr-qc/0012103}}].

\bibitem{Sachs:1962wk}
R.~K. Sachs, \emph{{Gravitational waves in general relativity. 8. Waves in
  asymptotically flat space-times}},
  \href{http://dx.doi.org/10.1098/rspa.1962.0206}{\emph{Proc. Roy. Soc. Lond.}
  {\bf A270} (1962) 103--126}.

\bibitem{Campoleoni:2016vsh}
A.~Campoleoni, H.~A. Gonzalez, B.~Oblak and M.~Riegler, \emph{{BMS Modules in
  Three Dimensions}}, \href{http://dx.doi.org/10.1142/S0217751X16500688,
  10.1142/9789813144101_0011}{\emph{Int. J. Mod. Phys.} {\bf A31} (2016)
  1650068}, [\href{https://arxiv.org/abs/1603.03812}{{\tt 1603.03812}}].

\bibitem{Hijano:2017eii}
E.~Hijano and C.~Rabideau, \emph{{Holographic entanglement and Poincaré blocks
  in three-dimensional flat space}},
  \href{http://dx.doi.org/10.1007/JHEP05(2018)068}{\emph{JHEP} {\bf 05} (2018)
  068}, [\href{https://arxiv.org/abs/1712.07131}{{\tt 1712.07131}}].

\bibitem{Hijano:2018nhq}
E.~Hijano, \emph{{Semi-classical BMS$_{3}$ blocks and flat holography}},
  \href{http://dx.doi.org/10.1007/JHEP10(2018)044}{\emph{JHEP} {\bf 10} (2018)
  044}, [\href{https://arxiv.org/abs/1805.00949}{{\tt 1805.00949}}].

\bibitem{Bagchi:2019unf}
A.~Bagchi, A.~Saha and Zodinmawia, \emph{{BMS Characters and Modular
  Invariance}},  \href{https://arxiv.org/abs/1902.07066}{{\tt 1902.07066}}.

\bibitem{Araujo:2018dem}
T.~Araujo, \emph{{Remarks on BMS3 invariant field theories: Correlation
  functions and nonunitary CFTs}},
  \href{http://dx.doi.org/10.1103/PhysRevD.98.026014}{\emph{Phys. Rev.} {\bf
  D98} (2018) 026014}, [\href{https://arxiv.org/abs/1802.06559}{{\tt
  1802.06559}}].

\bibitem{Witten:2007kt}
E.~Witten, \emph{{Three-Dimensional Gravity Revisited}},
  \href{https://arxiv.org/abs/0706.3359}{{\tt 0706.3359}}.

\bibitem{Castro:2011zq}
A.~Castro, M.~R. Gaberdiel, T.~Hartman, A.~Maloney and R.~Volpato, \emph{{The
  Gravity Dual of the Ising Model}},
  \href{http://dx.doi.org/10.1103/PhysRevD.85.024032}{\emph{Phys. Rev.} {\bf
  D85} (2012) 024032}, [\href{https://arxiv.org/abs/1111.1987}{{\tt
  1111.1987}}].

\bibitem{Barnich:2012rz}
G.~Barnich, A.~Gomberoff and H.~A. González, \emph{{Three-dimensional
  Bondi-Metzner-Sachs invariant two-dimensional field theories as the flat
  limit of Liouville theory}},
  \href{http://dx.doi.org/10.1103/PhysRevD.87.124032}{\emph{Phys. Rev.} {\bf
  D87} (2013) 124032}, [\href{https://arxiv.org/abs/1210.0731}{{\tt
  1210.0731}}].

\bibitem{Barnich:2013yka}
G.~Barnich and H.~A. Gonzalez, \emph{{Dual dynamics of three dimensional
  asymptotically flat Einstein gravity at null infinity}},
  \href{http://dx.doi.org/10.1007/JHEP05(2013)016}{\emph{JHEP} {\bf 05} (2013)
  016}, [\href{https://arxiv.org/abs/1303.1075}{{\tt 1303.1075}}].

\bibitem{Kabat:2011rz}
D.~Kabat, G.~Lifschytz and D.~A. Lowe, \emph{{Constructing local bulk
  observables in interacting AdS/CFT}},
  \href{http://dx.doi.org/10.1103/PhysRevD.83.106009}{\emph{Phys. Rev.} {\bf
  D83} (2011) 106009}, [\href{https://arxiv.org/abs/1102.2910}{{\tt
  1102.2910}}].

\bibitem{Kabat:2016zzr}
D.~Kabat and G.~Lifschytz, \emph{{Locality, bulk equations of motion and the
  conformal bootstrap}},
  \href{http://dx.doi.org/10.1007/JHEP10(2016)091}{\emph{JHEP} {\bf 10} (2016)
  091}, [\href{https://arxiv.org/abs/1603.06800}{{\tt 1603.06800}}].

\bibitem{Weinberg:1965nx}
S.~Weinberg, \emph{{Infrared photons and gravitons}},
  \href{http://dx.doi.org/10.1103/PhysRev.140.B516}{\emph{Phys. Rev.} {\bf 140}
  (1965) B516--B524}.

\bibitem{Carney:2018ygh}
D.~Carney, L.~Chaurette, D.~Neuenfeld and G.~Semenoff, \emph{{On the need for
  soft dressing}}, \href{http://dx.doi.org/10.1007/JHEP09(2018)121}{\emph{JHEP}
  {\bf 09} (2018) 121}, [\href{https://arxiv.org/abs/1803.02370}{{\tt
  1803.02370}}].

\bibitem{Carney:2017oxp}
D.~Carney, L.~Chaurette, D.~Neuenfeld and G.~W. Semenoff, \emph{{Dressed
  infrared quantum information}},
  \href{http://dx.doi.org/10.1103/PhysRevD.97.025007}{\emph{Phys. Rev.} {\bf
  D97} (2018) 025007}, [\href{https://arxiv.org/abs/1710.02531}{{\tt
  1710.02531}}].

\bibitem{Lewkowycz:2016ukf}
A.~Lewkowycz, G.~J. Turiaci and H.~Verlinde, \emph{{A CFT Perspective on
  Gravitational Dressing and Bulk Locality}},
  \href{http://dx.doi.org/10.1007/JHEP01(2017)004}{\emph{JHEP} {\bf 01} (2017)
  004}, [\href{https://arxiv.org/abs/1608.08977}{{\tt 1608.08977}}].

\bibitem{Anand:2017dav}
N.~Anand, H.~Chen, A.~L. Fitzpatrick, J.~Kaplan and D.~Li, \emph{{An Exact
  Operator That Knows Its Location}},
  \href{http://dx.doi.org/10.1007/JHEP02(2018)012}{\emph{JHEP} {\bf 02} (2018)
  012}, [\href{https://arxiv.org/abs/1708.04246}{{\tt 1708.04246}}].

\bibitem{Chen:2019hdv}
H.~Chen, J.~Kaplan and U.~Sharma, \emph{{AdS$_3$ Reconstruction with General
  Gravitational Dressings}},  \href{https://arxiv.org/abs/1905.00015}{{\tt
  1905.00015}}.

\bibitem{Cotler:2017erl}
J.~Cotler, P.~Hayden, G.~Penington, G.~Salton, B.~Swingle and M.~Walter,
  \emph{{Entanglement Wedge Reconstruction via Universal Recovery Channels}},
  \href{https://arxiv.org/abs/1704.05839}{{\tt 1704.05839}}.

\bibitem{Chen:2019gbt}
C.-F. Chen, G.~Penington and G.~Salton, \emph{{Entanglement Wedge
  Reconstruction using the Petz Map}},
  \href{https://arxiv.org/abs/1902.02844}{{\tt 1902.02844}}.

\bibitem{Higuchi:1986wu}
A.~Higuchi, \emph{{Symmetric Tensor Spherical Harmonics on the $N$ Sphere and
  Their Application to the De Sitter Group SO($N$,1)}},
  \href{http://dx.doi.org/10.1063/1.527513}{\emph{J. Math. Phys.} {\bf 28}
  (1987) 1553}.

\bibitem{Mack:2009mi}
G.~Mack, \emph{{D-independent representation of Conformal Field Theories in D
  dimensions via transformation to auxiliary Dual Resonance Models. Scalar
  amplitudes}},  \href{https://arxiv.org/abs/0907.2407}{{\tt 0907.2407}}.

\bibitem{Mack:2009gy}
G.~Mack, \emph{{D-dimensional Conformal Field Theories with anomalous
  dimensions as Dual Resonance Models}}, {\emph{Bulg. J. Phys.} {\bf 36} (2009)
  214--226}, [\href{https://arxiv.org/abs/0909.1024}{{\tt 0909.1024}}].

\bibitem{Barnich:2014kra}
G.~Barnich and B.~Oblak, \emph{{Notes on the BMS group in three dimensions: I.
  Induced representations}},
  \href{http://dx.doi.org/10.1007/JHEP06(2014)129}{\emph{JHEP} {\bf 06} (2014)
  129}, [\href{https://arxiv.org/abs/1403.5803}{{\tt 1403.5803}}].

\bibitem{Harlow:2011ke}
D.~Harlow and D.~Stanford, \emph{{Operator Dictionaries and Wave Functions in
  AdS/CFT and dS/CFT}},  \href{https://arxiv.org/abs/1104.2621}{{\tt
  1104.2621}}.

\bibitem{Bagchi:2017cpu}
A.~Bagchi, M.~Gary and Zodinmawia, \emph{{The nuts and bolts of the BMS
  Bootstrap}}, \href{http://dx.doi.org/10.1088/1361-6382/aa8003}{\emph{Class.
  Quant. Grav.} {\bf 34} (2017) 174002},
  [\href{https://arxiv.org/abs/1705.05890}{{\tt 1705.05890}}].

\bibitem{Moreira:1995he}
E.~S. Moreira, Jr., \emph{{Massive quantum fields in a conical background}},
  \href{http://dx.doi.org/10.1016/0550-3213(95)00357-X}{\emph{Nucl. Phys.} {\bf
  B451} (1995) 365--378}, [\href{https://arxiv.org/abs/hep-th/9502016}{{\tt
  hep-th/9502016}}].

\bibitem{Arefeva:2016wek}
I.~{\relax Ya}. Aref'eva and M.~A. Khramtsov, \emph{{AdS/CFT prescription for
  angle-deficit space and winding geodesics}},
  \href{http://dx.doi.org/10.1007/JHEP04(2016)121}{\emph{JHEP} {\bf 04} (2016)
  121}, [\href{https://arxiv.org/abs/1601.02008}{{\tt 1601.02008}}].

\bibitem{Hollands:2014eia}
S.~Hollands and R.~M. Wald, \emph{{Quantum fields in curved spacetime}},
  \href{http://dx.doi.org/10.1016/j.physrep.2015.02.001}{\emph{Phys. Rept.}
  {\bf 574} (2015) 1--35}, [\href{https://arxiv.org/abs/1401.2026}{{\tt
  1401.2026}}].

\bibitem{Vilenkin:1984ib}
A.~Vilenkin, \emph{{Cosmic Strings and Domain Walls}},
  \href{http://dx.doi.org/10.1016/0370-1573(85)90033-X}{\emph{Phys. Rept.} {\bf
  121} (1985) 263--315}.

\bibitem{Gott:1984ef}
J.~R. Gott, III, \emph{{Gravitational lensing effects of vacuum strings: Exact
  solutions}}, \href{http://dx.doi.org/10.1086/162808}{\emph{Astrophys. J.}
  {\bf 288} (1985) 422--427}.

\bibitem{Hijano:2015qja}
E.~Hijano, P.~Kraus, E.~Perlmutter and R.~Snively, \emph{{Semiclassical
  Virasoro blocks from AdS$_{3}$ gravity}},
  \href{http://dx.doi.org/10.1007/JHEP12(2015)077}{\emph{JHEP} {\bf 12} (2015)
  077}, [\href{https://arxiv.org/abs/1508.04987}{{\tt 1508.04987}}].

\bibitem{Maloney:2016kee}
A.~Maloney, H.~Maxfield and G.~S. Ng, \emph{{A conformal block Farey tail}},
  \href{http://dx.doi.org/10.1007/JHEP06(2017)117}{\emph{JHEP} {\bf 06} (2017)
  117}, [\href{https://arxiv.org/abs/1609.02165}{{\tt 1609.02165}}].

\bibitem{AUXEHCR}
E.~Hijano and C.~Rabideau, \emph{{to appear}}, .

\bibitem{He:2014laa}
T.~He, V.~Lysov, P.~Mitra and A.~Strominger, \emph{{BMS supertranslations and
  Weinberg’s soft graviton theorem}},
  \href{http://dx.doi.org/10.1007/JHEP05(2015)151}{\emph{JHEP} {\bf 05} (2015)
  151}, [\href{https://arxiv.org/abs/1401.7026}{{\tt 1401.7026}}].

\bibitem{Strominger:2014pwa}
A.~Strominger and A.~Zhiboedov, \emph{{Gravitational Memory, BMS
  Supertranslations and Soft Theorems}},
  \href{http://dx.doi.org/10.1007/JHEP01(2016)086}{\emph{JHEP} {\bf 01} (2016)
  086}, [\href{https://arxiv.org/abs/1411.5745}{{\tt 1411.5745}}].

\bibitem{Maldacena:2015iua}
J.~Maldacena, D.~Simmons-Duffin and A.~Zhiboedov, \emph{{Looking for a bulk
  point}}, \href{http://dx.doi.org/10.1007/JHEP01(2017)013}{\emph{JHEP} {\bf
  01} (2017) 013}, [\href{https://arxiv.org/abs/1509.03612}{{\tt 1509.03612}}].

\bibitem{ListIntegrals}
S.~Gradshteyn, I. and M.~Ryzhiz, I., \emph{{Table of integrals, series, and
  products}}, .

\end{thebibliography}\endgroup
\end{document}